\documentclass[aps,prb,longbibliography,groupedaddress,twocolumn,10pt]{revtex4-2}
\usepackage{graphicx}
\usepackage{mathtools}
\usepackage{amsmath}
\usepackage{amssymb}
\usepackage{bbm}
\usepackage{bm}
\usepackage{cancel}
\usepackage{xspace}
\usepackage{braket}
\usepackage{xcolor}
\usepackage{siunitx}
\usepackage{mathrsfs}
\usepackage{comment}
\usepackage{placeins}

\usepackage[export]{adjustbox}
\usepackage{wrapfig}
\usepackage{bbold}
\usepackage{booktabs} 

\usepackage{algpseudocode}

\usepackage[colorlinks=true]{hyperref} 

\newcommand{\ee}{\mathrm{e}}  
\DeclareMathOperator*{\ii}{i} 
\newcommand*\dd{\mathop{}\!\mathrm{d}}

\renewcommand{\vec}[1]{\bm{#1}} 
\newcommand{\mat}[1]{\bm{#1}} 
\newcommand{\kel}[1]{\underline{#1}} 

\definecolor{hblue}{RGB}{0,80,255}
\definecolor{hred}{RGB}{255,80,0}

\newif\ifdraftmode\draftmodetrue

\newcommand{\FEC}{fit-error corrected AMEA\xspace}
\newcommand{\fec}{fit-error correction\xspace}
\newcommand{\bratio}{Boltzmann ratio\xspace}
\newcommand{\Brat}{\mathcal{B}}
\newcommand{\Teff}{T_{\mathrm{eff}}}
\newcommand{\kk}{\vec{k}}

\newcommand{\Dphys}{\kel{\Delta}_{\mathrm{ph}}}
\newcommand{\Daux}{\kel{\Delta}_{\mathrm{aux}}}

\begin{document}

\title{Fit-error corrected auxiliary master equation approach \\ for
     low-temperature nonequilibrium dynamical mean-field theory}

\author{Sebastian Alois Fuchs}
\email[]{s.fuchs@tugraz.at}
\author{Enrico Arrigoni}
\email[]{arrigoni@tugraz.at}
\author{Tommaso Maria Mazzocchi}
\email[]{mazzocchi@tugraz.at}
\affiliation{Institute of Theoretical and Computational Physics, Graz University of Technology, 8010 Graz, Austria}

\date{\today}


\begin{abstract}

The auxiliary master equation approach (AMEA) is a nonperturbative impurity
solver for steady-state nonequilibrium dynamical mean-field theory (DMFT). It
replaces the physical hybridization function by that of a small auxiliary open
quantum system, whose parameters are fitted to reproduce it, and then solves
the resulting many-body problem essentially exactly. AMEA's accuracy is thus
limited by the quality of this fit: since the fastest-varying part of the
hybridization is set by the distribution function, whose Fermi edges sharpen
at low temperature, the fit tends to overestimate the temperature, especially
for weak reservoir coupling. For the auxiliary hybridization itself, however,
AMEA's self-energy is essentially exact.

We correct for the fit error by estimating, via an approximate solver
applicable to arbitrary hybridization functions, the ratio between the
self-energies of the physical and auxiliary hybridizations, and using it to
extend AMEA's self-energy from the auxiliary to the physical case. This
fit-error correction becomes exact whenever either the fit or the approximate
solver is exact.

We benchmark the scheme using the nonequilibrium Kajueter-Kotliar iterated
perturbation theory (IPT), generalized here to spin-polarized solutions, as
the approximate solver; other approximate solvers, such as the noncrossing
approximation, can be used as well. Applied to an Anderson impurity model
with a spin-polarized bath, and within DMFT to a Hubbard layer weakly coupled
to two metallic reservoirs, the correction largely removes the spurious
heating from the AMEA fit and from insufficient thermalization by the
reservoirs, while preserving the correct description of the Mott insulator,
which IPT alone cannot fully recover.
\end{abstract}

\maketitle

\section{Introduction}\label{sec:intro}

Correlated electron systems driven into a nonequilibrium steady state by a
current or a bias display a wealth of phenomena, ranging from the dielectric
breakdown and resistive switching of Mott insulators
\cite{ec.ok.10,ar.ko.12,ar.we.13,li.ar.15,ma.am.16,mu.we.18,ha.ar.23,di.ha.23,le.pr.12}
to the current-induced manipulation of magnetic order. The latter is the
operating principle of antiferromagnetic spintronics, in which the N\'eel
vector of metallic antiferromagnets such as CuMnAs and Mn$_2$Au is reoriented
by current pulses~\cite{wa.ho.16,ju.ma.16,ba.ma.18}, and it has received
renewed interest with the discovery of altermagnets~\cite{sm.si.22}.
Correlated antiferromagnetic Mott insulators respond to a current as well: in
Sr$_2$IrO$_4$ it lowers the N\'eel temperature~\cite{ca.te.18}, and in
Ca$_2$RuO$_4$ it drives an insulator-to-metal transition whose
current-carrying states can coexist with, or suppress, the antiferromagnetic
order~\cite{je.wi.20}. A question that accompanies all these experiments is
whether the observed effect is an intrinsic nonequilibrium phenomenon or
merely a consequence of Joule heating, and the answer is often far from
obvious~\cite{me.gr.18,ch.hu.19,ma.yo.20,je.wi.20,zi.ai.13,ka.ca.20}.

From the theoretical standpoint, a driven system reaches a stationary state
only if the injected energy is removed by a dissipative
environment~\cite{ha.li.13,am.we.12,ar.we.13}. The temperature of the
electrons or, more generally, their nonequilibrium distribution function is
then an \emph{outcome} of the balance between injection and dissipation rather
than an input parameter. Since broken-symmetry phases such as magnetic order
exist only below a characteristic temperature, whether they survive a drive
depends on how far this distribution departs from a cold Fermi
function~\cite{we.ts.12,ts.ec.13,ts.we.13}. A theory that aims to separate
intrinsic nonequilibrium effects from heating must therefore describe the
distribution function accurately at temperatures well below the ordering
scale, and it must not heat the electrons on its own.

Dynamical mean-field theory (DMFT)~\cite{me.vo.89,ge.ko.92,ge.ko.96,jarr.92}
and its nonequilibrium extensions~\cite{sc.mo.02u,fr.tu.06,ao.ts.14} offer a
natural framework for this problem. For steady states, the theory can be
formulated directly on the real-frequency axis without time propagation
\cite{jo.fr.08,okam.07,ar.kn.13,ti.do.15,so.do.18,ma.ga.22,ga.we.24}, which
gives immediate access to the current and to the spectral and distribution
functions. In this work we focus on such nonequilibrium steady states, with
equilibrium as a special case that serves as a benchmark. 

Within DMFT, the lattice problem is mapped onto a correlated impurity coupled
to a self-consistently determined environment,\footnote{Throughout this paper, the
terms bath, environment, reservoir, and lead are used synonymously for
noninteracting fermionic degrees of freedom coupled to the correlated system,
which can be integrated out and enter only via a hybridization function.
Within DMFT, the physical bath of the impurity consists of the rest of the
lattice together with the reservoirs attached to it.} characterized by its
\emph{hybridization function} $\kel{\Delta}$,\footnote{We denote two-by-two
matrices in Keldysh space by an underline, $\kel{\cdots}$, see, e.g.,
Ref.~\cite{ra.sm.86} and Eq.~\eqref{kelx} below.} which is itself out of
equilibrium. The task of the impurity solver is to determine the self-energy
$\kel{\Sigma}$ of this impurity problem, which, within DMFT, is identified
with the local self-energy of the lattice. Numerically exact approaches to this nonequilibrium impurity
problem exist, among them real-time and Keldysh quantum Monte
Carlo~\cite{we.ok.09,co.gu.15,pr.gr.15,be.pa.19}, scattering-states numerical
renormalization group~\cite{ande.08}, and tensor-network methods with
Lindblad-driven discretized leads~\cite{sc.go.16,sc.we.18,lo.we.20}.
However, their use inside a self-consistent DMFT lattice calculation has so
far remained the exception, mainly because of their computational cost, which
has to be paid at every iteration of the self-consistency. 

The auxiliary master equation approach (AMEA)
\cite{ar.kn.13,do.nu.14,do.ga.15,do.so.17,fu.do.18,we.lo.23} is a
non-perturbative impurity solver devised for this task.

It replaces the physical bath of the impurity by an \emph{auxiliary bath},
consisting of a few bath sites coupled to Markovian reservoirs. The
parameters of the auxiliary bath are fitted such that its hybridization
function, $\Daux$, reproduces the physical one, $\Dphys$, as closely as
possible. The dynamics of the resulting finite open quantum system is
governed by a Lindblad master equation for its many-body density matrix,
which can then be solved essentially exactly. This has been done by
Krylov-space (Lanczos) methods~\cite{do.nu.14}, matrix product
states~\cite{do.ga.15,fu.ba.20}, stochastic wave functions~\cite{so.fu.19},
or, in the most recent implementation, by a configuration-interaction (CI)
expansion~\cite{we.lo.23}. Thanks to the Markovian reservoirs, the fit accuracy
converges exponentially with the number of bath sites~\cite{do.so.17}. 

The method has been successfully applied to a variety
of nonequilibrium situations, see, e.g.,
Refs.~\cite{ma.ga.22,ga.ma.22,ma.we.23,ma.we.24,ma.we.25,ma.ai.26}.

The limitation of AMEA lies in the fit, whose accuracy degrades with decreasing temperatures. Out of equilibrium the hybridization
function has a retarded and a Keldysh component, and the latter is
essentially the product of the bath spectral function and of its
distribution function.
Even when the spectral part of $\Dphys$ is smooth, its
most rapidly varying part is thus set by the distribution function, whose
Fermi edges become sharper as the temperature is lowered. A finite Markovian
auxiliary bath necessarily smears these edges~\cite{sc.go.16,el.gr.17}, so
that the impurity effectively experiences a higher temperature than the
physical one. This sets a lower bound on the
temperatures that can be described reliably, in the same way as the level
spacing of the bath does for exact diagonalization in
equilibrium~\cite{ca.kr.94,ko.sa.08,sene.10,li.is.12,me.ze.20}.

The important point, however, is that this is not an error of the many-body
treatment: for the auxiliary hybridization function $\Daux$ resulting from
the fit, AMEA is exact, up to the CI truncation, which can be controlled. In
other words, AMEA provides the exact self-energy $\kel{\Sigma}[\Daux]$ of the
auxiliary hybridization function, where the square brackets denote the
dependence on the hybridization function, rather than the self-energy
$\kel{\Sigma}[\Dphys]$ of the physical one, which is the quantity of interest.

This suggests a natural strategy. If the relation between $\kel{\Sigma}[\Dphys]$
and $\kel{\Sigma}[\Daux]$ were known, the AMEA result could be extrapolated from the
auxiliary to the physical hybridization function. This relation is not accessible
to AMEA itself, but it can be estimated by an approximate solver which, in
contrast to AMEA, can be applied directly, i.e., without a fit, to
\emph{arbitrary} hybridization functions $\kel{\Delta}$, providing an
approximate self-energy $\kel{\Sigma}_{\mathrm{app}}[\kel{\Delta}]$.
Evaluating $\kel{\Sigma}_{\mathrm{app}}$ for both $\Dphys$ and $\Daux$ yields
an estimate of the ratio between $\kel{\Sigma}[\Dphys]$ and
$\kel{\Sigma}[\Daux]$, which is then applied to the AMEA result,
\begin{equation}
  \kel{\Sigma}[\Dphys]
  = \kel{\Sigma}[\Daux]\,\frac{\kel{\Sigma}[\Dphys]}{\kel{\Sigma}[\Daux]}
  \approx \kel{\Sigma}_{\mathrm{AMEA}}[\Daux]\,
     \frac{\kel{\Sigma}_{\mathrm{app}}[\Dphys]}{\kel{\Sigma}_{\mathrm{app}}[\Daux]}.
  \label{eq:fec_schematic}
\end{equation}
Equation~\eqref{eq:fec_schematic} is schematic and is specified in
Sec.~\ref{sec:mixing}: since the self-energy is a matrix in Keldysh space,
the ratio is applied separately to its spectral part and to the
\emph{\bratio} $F/(1-F)$ of its distribution function $F$. In equilibrium,
the latter is the Boltzmann factor $\ee^{-(\omega-\mu)/T}$, with $\omega$ the
frequency, $\mu$ the chemical potential, and $T$ the temperature, so that its correction amounts to
a correction of the temperature or, out of equilibrium, of an
\emph{effective temperature}. We refer to this strategy as \emph{\fec}.

The \fec becomes exact in two limits. (i) When the fit is perfect,
$\Daux = \Dphys$, AMEA is exact also for the physical problem, and the ratio
in Eq.~\eqref{eq:fec_schematic} is identically one, independently of the
approximate solver. The \fec then leaves the AMEA result unchanged, so that
its nonperturbative content is preserved. (ii) When the approximate solver
becomes exact,\footnote{For example, for weak interaction, or when higher
orders of perturbation theory are included.} the ratio is exact,
independently of the quality of the fit. While these two limits are ideal and
in practice never reached, except for trivial cases, the residual error is
typically small when one of the two errors is small. The approximate solver
thus does not need to provide an accurate self-energy, only an accurate
relative change, which is why the scheme remains useful even in regimes where
the approximate solver alone fails qualitatively.

Natural candidates for the approximate solver are perturbative impurity
solvers, which can be applied to an arbitrary hybridization function, without
any restriction on its form.
 In this work we use iterated perturbation theory (IPT), which
evaluates the second-order self-energy diagram on the bath Green's
function~\cite{ge.ko.96}. In the form of Kajueter and Kotliar
(KK-IPT)~\cite{ka.ko.96}, 
this contribution is interpolated towards the atomic
limit, which extends the method to arbitrary filling and lets it capture the
Hubbard bands. Steady-state versions of IPT have long been used in
nonequilibrium DMFT~\cite{sc.mo.02u,ar.ko.12,ts.we.13,ma.ar.24}, and KK-IPT
has recently been extended to nonequilibrium steady states at arbitrary
filling~\cite{ma.ar.26}. KK-IPT is an explicit functional of the
hybridization function, including its Keldysh component, and can thus be
evaluated for $\Dphys$ and $\Daux$ alike. In particular, for an equilibrium
$\Dphys$ the resulting self-energy inherits its distribution function, and
thus its temperature, exactly.
Being perturbative at its core, it is not quantitatively reliable for a strong
local interaction $U$, where it overestimates the quasiparticle weight and can miss the
Mott insulator altogether. As argued above, this is less relevant when it is
only used to estimate the ratio in Eq.~\eqref{eq:fec_schematic}.

Since the target applications are magnetically ordered systems, we generalize
the nonequilibrium KK-IPT of Ref.~\cite{ma.ar.26} to spin-polarized solutions.
We refer to the resulting combination of AMEA with KK-IPT as \FEC{} (FEC-AMEA).
The choice of KK-IPT is, however, by no means essential. Other approximate
solvers, such as the noncrossing approximation and its
extensions~\cite{wi.me.94,ec.we.10,ge.ki.25}, which expand in the
hybridization rather than in the interaction and are therefore complementary
to IPT at strong coupling, can be used in the same way.

The need for such a correction is greatest within DMFT. As
discussed above, already for a single impurity the smearing of the Fermi
edges by the auxiliary bath leads to an overestimate of the temperature.
In the benchmarks of Sec.~\ref{sec:singleshot},  the effective temperature
of the self-energy exceeds the bath temperature by up to a factor of two, and
by even more at the lowest temperature considered.
Within DMFT, this error is amplified at each iteration of the
self-consistency: each AMEA self-energy carries a slightly hotter
distribution function, which is fed back via the lattice Dyson equation into
the next hybridization function, so that the loop is drawn towards smooth,
i.e., hot, hybridization functions. This effect becomes particularly strong
when the external reservoirs, which fix the temperature, are weakly coupled
to the lattice, as in the benchmarks of Sec.~\ref{sec:dmft}, or when their spectrum is suppressed at the Fermi level,
since their thermalizing effect is then reduced.
This is problematic for magnetic order, since the resulting
effective temperature can become comparable to the N\'eel temperature, so
that magnetism would be artificially suppressed. Moreover, the spurious heating alone suppresses the
quasiparticle peak in a way that mimics the vicinity of the Mott transition.
The \fec cures this artifact, since it corrects the distribution function
of the self-energy so that the self-consistency converges to a sharper and colder hybridization
function.

Remedies for the finite-bath error have been proposed before. In equilibrium,
one can average over an ensemble of bath representations~\cite{gr.st.12} or
enlarge the bath in an optimized single-particle basis~\cite{lu.ho.14}. 
Both in and out of equilibrium, one can include corrections 
in the difference $\Dphys - \Daux$ perturbatively by dual
fermions~\cite{ch.co.19} or by a functional interpolation~\cite{we.ar.25}.
The \fec follows a similar philosophy, in that it starts from the
exact solution of the auxiliary problem and corrects for the difference
between the two hybridization functions. In contrast to the dual-fermion
approach, it does not require the two-particle vertex of the auxiliary
system, whose evaluation becomes prohibitive for a large number of bath
sites.

We benchmark FEC-AMEA in two steps. We start from  a single impurity in a spin-polarized
environment, which isolates the solvers from the self-consistency. In equilibrium, FEC-AMEA
follows the AMEA spectral function while recovering the temperature of the
reservoir.

In full DMFT for the two-dimensional (2D) Hubbard model weakly coupled to a
thermal reservoir,  
it removes the artificial heating of the
self-consistent AMEA solution,
while keeping the nonperturbative Hubbard bands, including the Mott
insulator that KK-IPT alone misses. This allows us to disentangle thermal from
correlation effects, such as a spurious pseudogap produced by heating.

The paper is organized as follows. Section~\ref{sec:background} summarizes
the known background, i.e., steady-state DMFT, the impurity model, and the
two impurity solvers, and Sec.~\ref{sec:mixing} introduces the \fec and its
implementation for AMEA and KK-IPT.
Section~\ref{sec:singleshot} contains the single-impurity benchmark for a
spin-polarized hybridization function, and Sec.~\ref{sec:dmft} the full DMFT
benchmark for the 2D lattice. We conclude in Sec.~\ref{sec:concl}.
Appendix~\ref{sec:dmft:mechanism} discusses the origin of the spurious
heating within DMFT, Appendix~\ref{sec:distfunc_grid} shows the distribution functions of the
single-impurity benchmark, Appendix~\ref{sec:spin_kkipt} derives the
spin-dependent nonequilibrium KK-IPT, and Appendix~\ref{app:sumrules} addresses
the spectral sum rules.


\section{Background: steady-state DMFT and impurity solvers}\label{sec:background}

This section summarizes the known ingredients of our approach, mainly to fix
the notation: steady-state DMFT (Sec.~\ref{sec:setup}), the Anderson impurity
model onto which it maps the lattice (Sec.~\ref{sec:solvers}), and the two
impurity solvers combined in this work, AMEA (Sec.~\ref{sec:solvers:amea})
and KK-IPT (Sec.~\ref{sec:solvers:ipt}). The only new element is the
generalization of KK-IPT to spin-polarized solutions, which is derived in
Appendix~\ref{sec:spin_kkipt}. Readers familiar with these methods may skip
directly to Sec.~\ref{sec:mixing}, where the fit-error correction is
introduced.

\subsection{Steady-state DMFT}\label{sec:setup}

To characterize the steady-state properties of the lattice at hand,
we use the steady-state formulation of nonequilibrium DMFT
\cite{ar.kn.13,ao.ts.14}, in which time-translational invariance is assumed from
the outset and all objects are functions of real frequency.

The only approximation underlying DMFT is that only {\em local correlation} in space
are retained, which allows to map the lattice onto a single-site impurity model coupled
to an environment to be determined self-consistently, the so-called {\em hybridization}
function~\cite{ge.ko.96}.

In this framework, a generic two-point function $\kel{X}$ is a $2\times2$ matrix in Keldysh space,
\begin{equation}
\label{kelx}
\kel{X} = \begin{pmatrix} X^{R} & X^{K} \\ 0 & X^{A}\end{pmatrix},
\end{equation}
with $X^{A} = (X^{R})^{\dagger}$ and $X^{R}$, $X^{K}$ and $X^{A}$ being the
{\em retarded}, {\em Keldysh} and {\em advanced} components~\cite{st.va.13}.

In this work, we focus on the single-band Hubbard model on a 2D square lattice coupled to noninteracting baths
\begin{align}
  \begin{split}
    \hat H & = -\sum_{\langle i,j \rangle\sigma} t_{ij}\, \hat c^{\dagger}_{i\sigma} \hat c^{\phantom{\dagger}}_{j\sigma}
            + U \sum_{i} \hat n_{i\uparrow} \hat n_{i\downarrow}
            + \sum_{i\sigma} \varepsilon_{\sigma} \hat n_{i\sigma} \\
            & + \hat H_{\text{bath}} + \hat H_{\text{coup}},
    \label{eq:H}
  \end{split}
\end{align}
where $\langle i, j\rangle$ denotes pairs of lattice sites and $t_{ij}$ are the matrix
elements of the hopping amplitudes in real space. Throughout this work we will
consider only nearest-neighbor hopping, i.e. $t_{ij} \equiv 1$ with $\mid \mat{r}_{i} - \mat{r}_{j} \mid = \pm 1$. 

The operator $\hat c^{\dagger}_{i\sigma}$ ($\hat c_{i\sigma}$) creates
(annihilates) an electron with spin $\sigma$ on lattice site $i$, and
$\hat n_{i\sigma} \equiv \hat c^{\dagger}_{i\sigma} \hat c_{i\sigma}$ is the particle
number operator, while $U \sum_{i} \hat n_{i\uparrow} \hat n_{i\downarrow}$ 
describes the local repulsion term.
The terms $\hat H_{\text{bath}}$ and $\hat H_{\text{coup}}$ in~\eqref{eq:H} account for
the electronic reservoir(s) and their (quadratic) coupling to the lattice.
Due to their non-interacting nature, their corresponding degrees of freedom
can be integrated out explicitly, and their effect on the lattice 
is encoded into the propagator $\kel{\gamma}_{\sigma}$, see Eq.~\eqref{eq:Glatt}, which for simplicity we take as being nondispersive (i.e. locally coupled to each site). Unless stated otherwise, they will be also  spin independent.

At the single-particle propagator level, the interaction between the electrons is encoded in the
 self-energy (SE), that within single-site DMFT 
is approximated to be 
{\em local}. In the present case, it
 can be spin-dependent, $\kel{\Sigma}_{\sigma}(\omega)$, so that magnetically ordered
solutions are allowed. Finding the SE is the task of the {\em impurity solver},
see Secs.~\ref{sec:solvers:amea} and \ref{sec:solvers:ipt}.

By applying a Fourier transform into the reciprocal space, the lattice Green's function (GF)
is obtained as
\begin{equation}
  {\kel{G}_{\sigma}(\kk,\omega)}^{-1} = {\kel{G}^{(0)}_{\sigma}(\kk,\omega)}^{-1}
    - \kel{\Sigma}_{\sigma}(\omega) - \kel{\gamma}_{\sigma}(\omega),
  \label{eq:Glatt}
\end{equation}
with $\kel{G}^{(0)}_{\sigma}(\kk,\omega)$ being the non-interacting GF of the 2D Hubbard lattice.\footnote{Its Keldysh component is in fact infinitesimal.}

The local electronic GF
\begin{equation}\label{eq:Glocal}
\kel{G}_{\sigma,\mathrm{loc}}(\omega) = \frac{1}{N}\sum_{\kk} \kel{G}_{\sigma}(\kk,\omega),
\end{equation}
with $N$ the number of lattice sites
determines the hybridisation function of the auxiliary impurity problem~\cite{ge.ko.96}
\begin{equation}
  \kel{\Delta}_{\mathrm{ph},\sigma}(\omega) = \kel{g}^{-1}_{0}(\omega)
    - \big[\kel{G}_{\mathrm{loc},\sigma}(\omega)\big]^{-1}
    - \kel{\Sigma}_{\sigma}(\omega)
  \label{eq:Delta}
\end{equation}
 via the Dyson equation and the
DMFT self-consistency condition $\kel{G}_{\sigma,\mathrm{loc}}(\omega) = \kel{G}_{\sigma,\mathrm{imp}}(\omega)$.
Here, $\kel{g}_{0}$ is the GF of an isolated lattice site, see
Eq.~\eqref{eq:G0imp} below.

\subsection{Anderson impurity model}\label{sec:solvers}

In this subsection we briefly introduce the nonequilibrium Anderson impurity
model (AIM) onto which the lattice problem is mapped within DMFT. A correlated  impurity is described by the Hamiltonian
\begin{equation}\label{eq:AIM}
\hat{H}_{\text{imp}} = \sum_{\sigma} \varepsilon_{\sigma} \, \hat{f}^{\dagger}_{\sigma} \hat{f}_{\sigma} + U \hat{n}_{\uparrow} \hat{n}_{\downarrow},
\end{equation}
with $\varepsilon_{\sigma}$ being a spin-dependent {\em onsite energy}.
The impurity is coupled to a non-interacting bath. Since the bath is
quadratic, its degrees of freedom can be integrated out, and its whole effect
on the impurity is encoded in the hybridization function
$\kel{\Delta}_{\sigma}(\omega)$. Out of equilibrium this is a Keldysh matrix of the
form~\eqref{kelx}.  In the present case, it is diagonal in spin since we do not consider
spin-flip processes. The noninteracting impurity GF thus reads
\begin{equation}\label{eq:G0imp}
\kel{G}^{-1}_{0,\sigma}(\omega) = \kel{g}^{-1}_{0,\sigma}(\omega) - \kel{\Delta}_{\sigma}(\omega),
\end{equation}
where $[\kel{g}^{-1}_{0,\sigma}]^{R}(\omega) = \omega - \varepsilon_{\sigma}$
is the inverse retarded GF of the isolated impurity, whose Keldysh component is
infinitesimal, and the interacting one is obtained from the Dyson equation
$\kel{G}^{-1}_{\sigma} = \kel{G}^{-1}_{0,\sigma} - \kel{\Sigma}_{\sigma}$.

The retarded component of $\kel{\Delta}_{\sigma}$ describes the spectrum of
the bath and its coupling to the impurity, while the Keldysh component
additionally encodes how the bath states are occupied,
\begin{equation}\label{eq:DeltaK}
\Delta^{K}_{\sigma}(\omega) = 2\ii\big[1-2F_{\Delta,\sigma}(\omega)\big]\,
    \mathrm{Im}\,\Delta^{R}_{\sigma}(\omega),
\end{equation}
with $F_{\Delta,\sigma}(\omega)$ the distribution function of the bath. In
equilibrium, $F_{\Delta,\sigma}$ is a Fermi function at the temperature and
chemical potential of the bath. Out of equilibrium, neither a temperature nor
a chemical potential can be defined in general, and $F_{\Delta,\sigma}$ is an
arbitrary function bounded between $0$ and $1$. For instance, for a bath
consisting of two leads held at different chemical potentials it is a
weighted average of their Fermi functions, and within DMFT it is determined
by the self-consistency~\eqref{eq:Delta}. A spin dependence of
$\kel{\Delta}_{\sigma}$ can originate from a spin-dependent coupling to the
bath or from a spin-polarized bath, e.g., a ferromagnetic contact.
The AIM is thus fully specified by $\varepsilon_{\sigma}$, $U$, and
$\kel{\Delta}_{\sigma}(\omega)$.
In the next subsections we will give a quick overview of the two techniques
used to solve the impurity model.

\subsection{Auxiliary master equation approach}\label{sec:solvers:amea}

In AMEA \cite{ar.kn.13,do.nu.14,do.so.17} the impurity problem~\eqref{eq:AIM},
defined by $\Dphys(\omega)$, is replaced by an \emph{open} quantum system
consisting of the impurity plus a small number $N_{B}$ of bath sites,
whose dynamics is governed by a Lindblad equation
\begin{align}
  \dot\rho = \mathcal{L}\rho = &-\ii[\hat H_{\mathrm{aux}},\rho]
  + \sum_{ij\sigma}\Gamma^{(1)}_{\sigma,ij}\Big(c_{j\sigma}\rho c^{\dagger}_{i\sigma}
  - \tfrac12\{c^{\dagger}_{i\sigma}c_{j\sigma},\rho\}\Big) \nonumber \\
  &+ \sum_{ij\sigma}\Gamma^{(2)}_{\sigma,ij}\Big(c^{\dagger}_{i\sigma}\rho c_{j\sigma}
  - \tfrac12\{c_{j\sigma}c^{\dagger}_{i\sigma},\rho\}\Big),
  \label{eq:lindblad}
\end{align}
with $\hat H_{\mathrm{aux}} = \sum_{ij\sigma} E_{\sigma,ij} c^{\dagger}_{i\sigma}
c_{j\sigma} + U n_{f\uparrow}n_{f\downarrow}$.

The parameters $\mat{E}_{\sigma}$,
$\mat{\Gamma}^{(1)}_{\sigma}$, $\mat{\Gamma}^{(2)}_{\sigma}$ are used to construct
the auxiliary hybridization $\Daux$ as an approximation to $\Dphys$, which is optimized in order to minimize the cost
function~\cite{do.nu.14,we.lo.23}
\begin{equation}
  \chi = \sum_\sigma\sum_{\alpha=R,K}\int\! \dd\omega\, W^{\alpha}(\omega)\,
   \Big[\mathrm{Im}\,\Delta^{\alpha}_{\mathrm{ph},\sigma}(\omega)
      - \mathrm{Im}\,\Delta^{\alpha}_{\mathrm{aux},\sigma}(\omega)\Big]^{2} ,
  \label{eq:cost}
\end{equation}
where $W^{\alpha}(\omega)$ is a weight function. Since the parameters of the
auxiliary system do not mix the two spin species, $\chi$ can be minimized
separately for each $\sigma$.
The error $\chi$ decreases exponentially as a function of $N_{B}$~\cite{do.so.17}.
The many-body Lindblad problem~\eqref{eq:lindblad} is solved 
by a Lanczos  treatment of the many-body density matrix with
a configuration interaction (CI) expansion around a
Hartree-Fock reference state~\cite{we.lo.23}~\footnote{The CI excitation cutoff is set to $N_{\mathrm{exc}}=3$ for the
lattice calculations of Sec.~\ref{sec:dmft} and
$N_{\mathrm{exc}}=4$ for the impurity solver benchmarks of
Sec.~\ref{sec:singleshot}. The larger cutoff is needed for the strongly
spin-polarized impurity of Sec.~\ref{sec:singleshot}, for which
$N_{\mathrm{exc}}=3$ can produce a negative spectral weight in the weak
Hubbard satellite of the majority spin. The paramagnetic lattice solutions of
Sec.~\ref{sec:dmft} are not affected by this problem, and we checked that they
show no such causality violation.}

To conclude, we want to emphasize that AMEA provides an \emph{exact} scheme 
for the auxiliary problem (up to the CI truncation),
i.e.\ it returns $\kel{\Sigma}[\Daux]$. Thus its
error is then inherited from the fit \eqref{eq:cost}.
Since the auxiliary bath is Markovian, a
finite $N_{B}$ cannot reproduce a sharp step in $\Delta^{K}$; the auxiliary bath
distribution is always somewhat smeared, which is equivalent to an artificially
increased temperature.

\subsection{Nonequilibrium KK-IPT}\label{sec:solvers:ipt}

The approximate solver we use is the nonequilibrium steady-state
extension~\cite{ma.ar.26} of the Kajueter-Kotliar iterated perturbation
theory~\cite{ka.ko.96}.  This approach builds a Weiss field $\mathscr{G}_{0\sigma}$
from the physical hybridization, evaluates the second-order diagram
$\tilde{\kel{\Sigma}}_{\sigma}$ on it, and interpolates that diagram towards
the atomic limit~\footnote{The auxiliary chemical potentials $\mu_{0\sigma}$
are fixed by requiring $n_{0\sigma} = n_{\sigma}$; the Keldysh component
follows from the same interpolation, i.e. no fluctuation-dissipation
theorem is employed to compute the various quantities in the Keldysh sector.}
with two coefficients $\alpha_{\sigma}$, $\beta_{\sigma}$ obtained by matching
the atomic-limit SE~\cite{ka.ko.96,po.he.98}. For details, see Appendix~\ref{sec:spin_kkipt}.

In our previous work, Ref.~\cite{ma.ar.26}, we formulated and benchmarked
this nonequilibrium version of the KK-IPT in the paramagnetic case,
$n_{\uparrow} = n_{\downarrow}$, whereas the settings the
present scheme is designed for also allow for magnetic polarizations. 
The spin-resolved generalization of the nonequilibrium KK-IPT is derived in
its entirety in Appendix~\ref{sec:spin_kkipt}. The appendix also gives access
to the spin-resolved spectral moment sum rules of which the third is
a genuinely cross-flavor, two-particle test and therefore the sharp one for
magnetic solutions and a Potthoff-Wegner-Nolting band-shift-type
correction~\cite{po.we.97} that restores it exactly. 

The key consideration is that, within IPT, the electronic SE is
an \emph{explicit functional} $\kel{\Sigma}_{\mathrm{IPT}}[\kel{\Delta}]$ that can be evaluated for \emph{any} hybridization function at negligible cost (a few
Fourier transforms), thus in equilibrium inheriting the distribution function of its input exactly.

\section{Fit-error corrected AMEA}\label{sec:mixing}

The \fec is defined schematically (but only schematically, for details see below) by Eq.~\eqref{eq:fec_schematic}: the SE
$\kel{\Sigma}_{\mathrm{AMEA}}[\Daux]$ obtained by AMEA for the auxiliary
hybridization function $\Daux$ of the fitted open system,
Eq.~\eqref{eq:cost}, is multiplied by the ratio between the SEs that an
approximate solver yields for the physical hybridization function $\Dphys$,
Eq.~\eqref{eq:Delta}, and for $\Daux$. As discussed in
Sec.~\ref{sec:solvers:amea}, AMEA is exact for $\Daux$, up to the CI
truncation, so that the only task of the approximate solver is to provide
the change of the SE from $\Daux$ to $\Dphys$, which is not accessible to
AMEA itself.

Equation~\eqref{eq:fec_schematic} becomes exact in two limits. (i) When the
fit is perfect, $\Daux = \Dphys$, the ratio is identically one for any
approximate solver, and the AMEA result, including its nonperturbative
content, is left unchanged. (ii) When the approximate solver is exact, the
ratio is exact, whatever the quality of the fit. Away from these limits, the
residual error is, to leading order, proportional to the product of the fit
error and of the error that the approximate solver makes in the relative
change of the SE. The approximate solver therefore does not need to
reproduce the SE itself accurately.

In this work we use KK-IPT as approximate solver and refer to the resulting
scheme as \FEC{} (FEC-AMEA). The additional cost is two KK-IPT solutions per
AMEA solution, i.e.\ a few percent of the total runtime.

Since the SE is a matrix in Keldysh space, Eq.~\eqref{eq:fec_schematic} has
to be specified for its components. The SE is fully determined by the
imaginary part of its retarded component, $\mathrm{Im}\,\Sigma^{R}$, from
which the real part follows by the Kramers-Kronig relation, and by its
distribution function $F_{\Sigma}$, which fixes the Keldysh component. We
therefore apply Eq.~\eqref{eq:fec_schematic} only to
$\mathrm{Im}\,\Sigma^{R}$ and to the \bratio $F_{\Sigma}/(1-F_{\Sigma})$, which both have
a fixed sign, rather than to the Keldysh component, which changes sign at the
Fermi level.

\emph{(i) Retarded component.} Applied to the imaginary part of the retarded
component, Eq.~\eqref{eq:fec_schematic} reads
\begin{equation}
  \mathrm{Im}\,\Sigma^{R}_{\mathrm{FEC},\sigma}(\omega)
  = \mathrm{Im}\,\Sigma^{R}_{\mathrm{AMEA},\sigma}(\omega) \;
    \frac{\mathrm{Im}\,\Sigma^{R}_{\mathrm{IPT},\sigma}(\omega)}
         {\mathrm{Im}\,\Sigma^{R}_{\mathrm{IPT\text{-}aux},\sigma}(\omega)} ,
  \label{eq:mixR}
\end{equation}
where $\Sigma_{\mathrm{IPT}}$ and $\Sigma_{\mathrm{IPT\text{-}aux}}$ denote the
KK-IPT solutions obtained with $\Dphys$ and with $\Daux$, respectively.  This
form manifestly preserves the sign of $\mathrm{Im}\,\Sigma^{R}$ and hence
causality. The real part is then reconstructed from the Kramers--Kronig
relation together with the exact Hartree shift,
\begin{equation}
  \mathrm{Re}\,\Sigma^{R}_{\mathrm{FEC},\sigma}(\omega) = U n_{\bar\sigma}
   + \frac{1}{\pi}\,\mathcal{P}\!\!\int\!\dd\omega'\,
     \frac{\mathrm{Im}\,\Sigma^{R}_{\mathrm{FEC},\sigma}(\omega')}{\omega'-\omega} ,
  \label{eq:mixKK}
\end{equation}
which guarantees that the corrected SE is analytic in the upper half
plane and has the correct first moment.

\emph{(ii) Distribution function.} The distribution function of the SE is
defined by
\begin{equation}
  F_{\Sigma}(\omega) = \frac12\left(1 - \frac{\mathrm{Im}\,\Sigma^{K}(\omega)}
                                         {2\,\mathrm{Im}\,\Sigma^{R}(\omega)}\right).
  \label{eq:Fdef}
\end{equation}
Since $F_{\Sigma}$ is bounded to $[0,1]$, a ratio applied to $F_{\Sigma}$
itself could push it outside this interval. We therefore use the
\bratio
\begin{equation}
  \Brat_{\Sigma}(\omega) = \frac{F_{\Sigma}(\omega)}{1-F_{\Sigma}(\omega)} ,
  \label{eq:Bdef}
\end{equation}
i.e., the ratio between the occupation and the vacancy of the states at
frequency $\omega$. Since $\Brat_{\Sigma}$ takes only positive values, it can
be corrected multiplicatively, in exactly the same way as
$\mathrm{Im}\,\Sigma^{R}$ in Eq.~\eqref{eq:mixR},
\begin{eqnarray}
  \Brat_{\mathrm{FEC},\sigma}(\omega) &=& \Brat_{\mathrm{AMEA},\sigma}(\omega) \;
    \frac{\Brat_{\mathrm{IPT},\sigma}(\omega)}{\Brat_{\mathrm{IPT\text{-}aux},\sigma}(\omega)} ,
\nonumber \\
  F_{\mathrm{FEC},\sigma} &=& \frac{\Brat_{\mathrm{FEC},\sigma}}{1+\Brat_{\mathrm{FEC},\sigma}} ,
  \label{eq:mixF}
\end{eqnarray}
which guarantees that $F_{\mathrm{FEC},\sigma}$ remains within $[0,1]$.~\footnote{For
the numerical implementation we recommend evaluating Eq.~\eqref{eq:mixF}
through the logarithm of $\Brat_{\Sigma}$ rather than the ratio itself:
$\ln\Brat_{\mathrm{FEC},\sigma} = \ln\Brat_{\mathrm{AMEA},\sigma}
+ \ln\Brat_{\mathrm{IPT},\sigma} - \ln\Brat_{\mathrm{IPT\text{-}aux},\sigma}$,
followed by $F_{\mathrm{FEC},\sigma} = \big[1+\exp(-\ln\Brat_{\mathrm{FEC},\sigma})\big]^{-1}$.
This is mathematically identical to Eq.~\eqref{eq:mixF} but numerically more
robust: it avoids the $0/0$ (or $\infty/\infty$) indeterminacy that direct
evaluation of the ratio produces whenever one of the three $F_{\Sigma}$'s
saturates to exactly $0$ or $1$, which happens away from the Fermi level
wherever $\mathrm{Im}\,\Sigma^{R}$ becomes small.} Moreover, since
$\Brat_{\Sigma} \to 1/\Brat_{\Sigma}$ under $F_{\Sigma} \to 1-F_{\Sigma}$,
particles and holes are treated on an equal footing. The corrected Keldysh
component of the SE is then recovered as
\begin{equation}
  \Sigma^{K}_{\mathrm{FEC},\sigma}(\omega) = 2\ii\big[1-2F_{\mathrm{FEC},\sigma}(\omega)\big]\,
    \mathrm{Im}\,\Sigma^{R}_{\mathrm{FEC},\sigma}(\omega).
  \label{eq:mixK}
\end{equation}
The \bratio is not merely a convenient parametrization. In equilibrium
$F_{\Sigma}$ is a Fermi function, and $\Brat_{\Sigma}$ reduces to the
Boltzmann factor
\begin{equation}
  \Brat_{\Sigma}(\omega) = \ee^{-(\omega-\mu)/T} ,
  \label{eq:Bequil}
\end{equation}
which motivates its name. A multiplicative correction of $\Brat_{\Sigma}$ is
thus precisely a redefinition of the effective temperature and of the
chemical potential, which is exactly the quantity that the fitting procedure
tends to overestimate. 

Out of equilibrium, the logarithmic derivative of
$\Brat_{\Sigma}$ generalizes this to a frequency-dependent inverse effective
temperature, and the multiplicative correction remains well defined. We then
introduce
\begin{equation}
  \Teff^{-1} = -\left.\frac{\partial \ln \Brat_{\Sigma}}{\partial\omega}\right|_{\omega=\mu}
  \label{eq:Teff}
\end{equation}
as the effective temperature used throughout this work, which, where
$F_{\Sigma}(\mu) = 1/2$, is equivalent to
$\Teff = -1/[4F_{\Sigma}'(\omega)]_{\omega=\mu}$.

Finally, Eqs.~\eqref{eq:mixR}-\eqref{eq:mixK} are applied to each spin
independently, and the fit-error corrected SE $\Sigma_{\mathrm{FEC}}$ is what enters
the DMFT self-consistency~\eqref{eq:Glatt}. In this way, both the spectral
part and the distribution function of the AMEA SE are extrapolated from the
auxiliary to the physical hybridization function, with the relative change
provided by KK-IPT.

\begin{figure}[t]
  \centering
  \includegraphics[width=\linewidth]{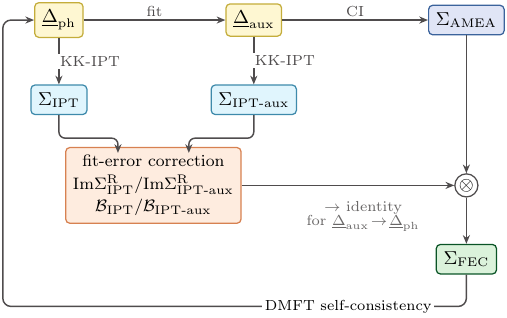}
  \caption{\label{fig:scheme} Construction of the fit-error corrected
  SE, Sec.~\ref{sec:mixing}. The physical hybridization $\Dphys$
  provided by the DMFT self-consistency, Eq.~\eqref{eq:Delta}, is represented by the hybridization
  $\Daux$ of a finite auxiliary open system [Eq.~\eqref{eq:cost}], which
  constitutes the input to the CI solver, giving $\Sigma_{\mathrm{AMEA}}$.
  KK-IPT is applied to \emph{both} hybridizations; the ratio of the
  imaginary parts of the two KK-IPT retarded self-energies, Eq.~\eqref{eq:mixR},
  and the corresponding ratio of their \bratio{}s
  $\Brat_{\Sigma} = F_{\Sigma}/(1-F_{\Sigma})$, Eq.~\eqref{eq:mixF},
  constitute the correction,
  which is applied to $\Sigma_{\mathrm{AMEA}}$ and yields the
  $\Sigma_{\mathrm{FEC}}$ that closes the DMFT loop.  Since both KK-IPT solvers
  use the same approximation of the same functional, the correction reduces to
  the identity when the fit is exact, $\Daux \to \Dphys$, so that no double
  counting is introduced.}
\end{figure}

\section{Results}

In order to benchmark the FEC-AMEA, we will first address the single-band AIM
with a spin-polarized hybridization function. Then we will investigate
the case of the paramagnetic Hubbard model on a 2D square lattice.
Magnetic lattice models will be addressed in future work. 

\subsection{Spin-polarized hybridization in the AIM}\label{sec:singleshot}

In order to mimic a {\em spin-polarized}, magnetic impurity, with majority and
minority spin species "experiencing" the spectral weight on opposite regions
of the chemical potential, we use a hybridization function
consisting of two Lorentzians displaced in opposite directions,
\begin{equation}
  \mathrm{Im}\,\Delta^{R}_{\mathrm{ph}, \sigma}(\omega) =
   -\frac{\Gamma^{2}}{(\omega - \sigma\,\omega_{0})^{2} + \Gamma^{2}} ,
  \label{eq:lorentz}
\end{equation}
with $\omega_{0} = \Gamma = 1$ and $\sigma = \pm 1$. In this equilibrium
setting, the spin-dependent {\em Keldysh components} of the hybridization 
are obtained by the fluctuation-dissipation theorem at temperature $T$.
For the benchmarks discussed in this section, $\Gamma$ sets the unit of energy.

\begin{figure*}[t]
  \centering
  \includegraphics[width=\textwidth]{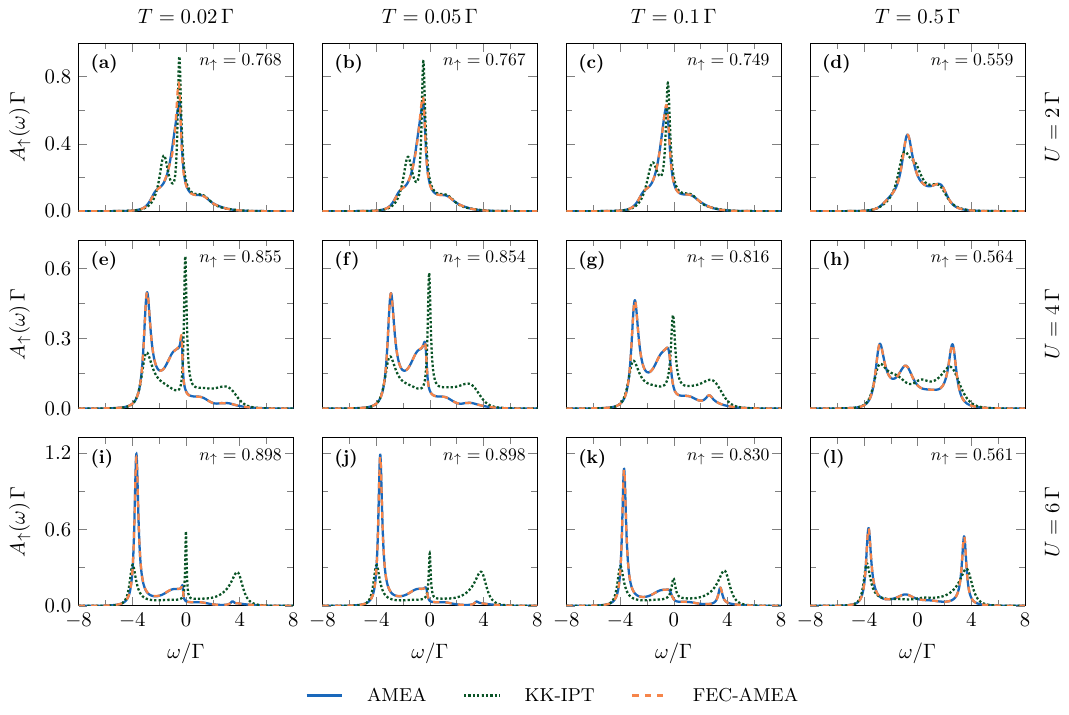}
\caption{\label{fig:singleshot} Comparison of the solutions to the AIM obtained with
  plain AMEA, KK-IPT, and FEC-AMEA for the spin-polarized hybridization
  function of Eq.~\eqref{eq:lorentz} with $\omega_0 = \Gamma$ and
  $\varepsilon_{\sigma} = -U/2$, in equilibrium at bath temperature $T$
  (columns), for different interaction strengths $U$ (rows).
  Only the majority species is shown because of
  the symmetry $A_{\downarrow}(\omega) = A_{\uparrow}(-\omega)$.
  The spectral functions $A_{\sigma}(\omega) = -\mathrm{Im}\,G^{R}_{\sigma}(\omega)/\pi$
  are obtained by inserting the SE of each solver into the impurity GF with
  the \emph{physical} hybridization, Eq.~\eqref{eq:G0imp}, so as to compare
  the three schemes on the same footing. The occupation of the majority species reported in each panel is obtained with AMEA.
  The \FEC{} curve is indistinguishable from AMEA on the scale of the plot,
  whereas KK-IPT retains a spurious quasiparticle peak at $\omega=0$
  whose spectral weight grows larger as $U$ is increased. 
AMEA uses $N_{B} = 6$ bath sites and $N_{\mathrm{exc}} = 4$.}
\end{figure*}

\begin{figure}[t]
  \centering
  \includegraphics[width=\linewidth]{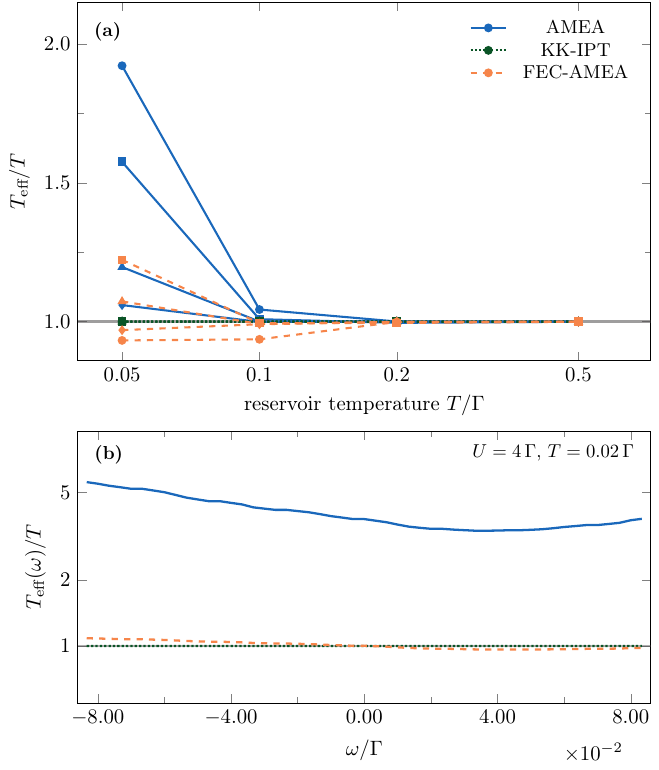}
\caption{\label{fig:teff_ss} Effective temperature of the impurity SE,
  Eq.~\eqref{eq:Teff}, divided by the reservoir temperature $T$, for plain AMEA,
  KK-IPT, and FEC-AMEA. The grey horizontal lines represent $\Teff = T$.
  (a) shows $\Teff/T$ obtained from the logarithmic slope of the \bratio
  $\Brat_{\Sigma}$ at $\omega = 0$, as a function of $T$ for $U = 2$
  (circles), $4$ (squares), $6$ (triangles), and $8\,\Gamma$ (diamonds).
  (b) Frequency-resolved $\Teff(\omega) =
  -[\partial_{\omega}\ln\Brat_{\Sigma}(\omega)]^{-1}$ for $T = 0.02\,\Gamma$
  and $U = 4\,\Gamma$ limited to the energy window $|\omega|/\Gamma < 0.08$.
  Other relevant parameters are as in Fig.~\ref{fig:singleshot}.}
\end{figure}

The results discussed below are obtained by fixing the hybridization profile,
Eq.~\eqref{eq:lorentz}, and varying the temperature $T$ and the onsite
interaction $U$. The resulting impurity spectral functions are shown
in Fig.~\ref{fig:singleshot}.

At $U = 2\,\Gamma$, the differences between KK-IPT, a genuinely perturbative
scheme, and plain AMEA are hardly noticeable on the scale of the plot almost
independently of the temperature, see Figs.~\ref{fig:singleshot}(a)-(d).
Notably, the results of the FEC-AMEA scheme also closely follow those obtained
with plain AMEA despite the (minimal) deviations introduced by IPT.

Despite the fact that the differences between KK-IPT and AMEA become more visible
as the interaction strength $U$ is increased, especially at low temperatures, the
FEC-AMEA solutions still closely agree with those obtained with plain AMEA
for larger values of the onsite interaction, see Figs.~\ref{fig:singleshot}(e)-(h)
and~\ref{fig:singleshot}(i)-(l). 

In particular, the most important difference is that KK-IPT retains a
pronounced (spurious) quasi-particle peak (QPP) at the chemical potential ($\omega = 0$)
whose spectral weight exceeds that of AMEA by ever larger
amounts as U is increased, while AMEA tends to suppress it together with the
spectral weight on the unoccupied part of the spectrum.

A tendency of IPT to overestimate the stability of the QPP is known already at half filling,
where, within DMFT, IPT yields a critical interaction for the Mott transition larger than that
obtained with numerically exact methods~\cite{bull.99,ge.ko.96}. Since within DMFT the
lattice enters only through the hybridization function, this is a property of IPT as an impurity
solver. For a given hybridization, IPT retains a QPP where the exact solution
of the corresponding AIM has already lost it. For a recent study on the persistence
of the QPP in a half-filled lattice in infinite dimensions both in and out of 
equilibrium, we point the reader to Ref.~\cite{ma.ar.24}, and corresponding discussion therein.

Away from half filling, the low-energy part of the KK-IPT spectrum is moreover largely
determined by the condition(s) used to fix the auxiliary chemical potential $\mu_{0\sigma}$,
rather than by the dynamics of the impurity~\cite{po.we.97}, which can favor the persistence
of a QPP also where, as in the present case, the spin-polarized environment suppresses it.

Also the SE obtained within the FEC-AMEA (not shown)
follows the solution of plain AMEA almost exactly, as it is expected
since the retarded correction~\eqref{eq:mixR} is proportional to the \emph{ratio} of two
KK-IPT solutions that differ only through the fit error, which ensures that the FEC-AMEA
result needs to approach unity wherever the fit performs well.

The correction introduced by the FEC-AMEA scheme is best appreciated from the
generalized distribution function, as spurious heating effects affect $F_{\Sigma}$ the most.
However, in order to keep the discussion in this section as simple as possible, the
analysis of the distribution function $F_{\Sigma}$ will be postponed to
Appendix~\ref{sec:distfunc_grid}.

Figure~\ref{fig:teff_ss}(a) shows the effective temperature estimated from the SE's distribution
function $F_{\Sigma}$, see Eq.~\eqref{eq:Teff}, for selected values of the interaction strength $U$.

For bath temperatures $T \ge 0.2\,\Gamma$ both plain AMEA and KK-IPT solvers reproduce
the bath temperature almost exactly, regardless of the interaction strength $U$. Below that
temperature, however, plain AMEA starts to show spurious heating effects. For instance,
at $T = 0.1\,\Gamma$ AMEA returns a $\Teff$ that is already too high, though by at most
$4\,\%$, across the range of $U$ values.

For even lower temperatures, $\Teff$ estimated from plain AMEA grows even larger.
For instance, at $T = 0.05\,\Gamma$, $\Teff$ exceeds $T$ by up to a factor 2.
Notably, the largest deviations occur at small values of $U$.

This behavior has been anticipated in Sec.~\ref{sec:solvers:amea}, where we pointed out
that an auxiliary system of finite size cannot represent a {\em step} in $\Delta^{K}$
that is sharper than its own broadening, so AMEA's resolved distribution is often smeared,
corresponding to a higher temperature solution.

On the other hand, KK-IPT inherits the distribution of its input
hybridization exactly, hence the temperature of the resulting KK-IPT resolved SE is exact
to within $1\,\%$ over the whole range of $U$ and $T$ used in our analysis.
The FEC-AMEA scheme therefore recovers the correct temperature or at least it limits
the spurious effect to only a few percent. For instance, at $T = 0.05\,\Gamma$, 
the error in $\Teff/T$ estimated with FEC-AMEA is reduced from a factor of two
with respect to plain AMEA, and at $T = 0.1\,\Gamma$ to below $6\,\%$,
see again Fig.~\ref{fig:teff_ss}(a).

For more results obtained with the FEC-AMEA scheme, we refer to Appendix~\ref{sec:distfunc_grid},
where we discuss the effect of the proposed scheme on the distribution function of the SE
$F_{\Sigma}(\omega)$ corresponding to the settings in Fig.~\ref{fig:singleshot}.

In the energy region where Eq.~\eqref{eq:Teff} is well defined, i.e. in the vicinity of the Fermi level
$\omega = 0$, we could also define an energy-dependent $\Teff$, by evaluating
the logarithmic derivative of $\Brat_{\Sigma}(\omega)$ at each frequency
rather than only at the chemical potential. However, one such definition makes sense
only if $F_{\Sigma}(\omega)$ is monotonous across the energy window of interest.

For plain AMEA and interaction strengths $U \ge 5\,\Gamma$, $F_{\Sigma}(\omega)$ shows
significant deviations from a monotonously evolving function, see Appendix~\ref{sec:distfunc_grid},
thus an effective temperature in the sense specified above cannot be used.

Figure~\ref{fig:teff_ss}(b) shows this frequency-resolved effective temperature
for the exemplary case with $U/\Gamma = 4$ and $T/\Gamma = 0.02$,
in the {\em thermal window} where $F_{\Sigma}(\omega)$ is well-behaved.
In this setting, $\Teff$ estimated via the FEC-AMEA scheme is flat throughout
the resolved energy window, and approaches the bath temperature while plain AMEA
largely overestimates $\Teff$.

We have shown that the FEC-AMEA scheme inherits the
spectral features of plain AMEA and in equilibrium the distribution function of KK-IPT, thus
retaining the best of both approaches at the same time.

As already mentioned in Sec.~\ref{sec:intro}, the approximate scheme used in the
fit-error correction need not capture the qualitative behavior of the solution necessarily,
but rather provide a reasonable estimate of the difference between the
physical and auxiliary hybridizations.

\begin{figure}[t]
  \centering
  \includegraphics[width=\linewidth]{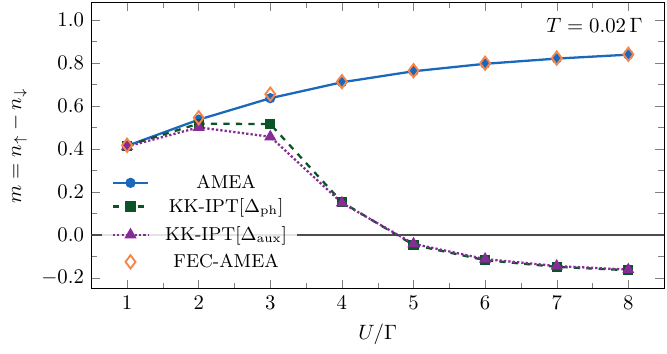}
\caption{\label{fig:breakdown}
  (a) Impurity magnetization $m = n_{\uparrow}-n_{\downarrow}$ as a function
  of $U$ at $T = 0.02\,\Gamma$, for plain AMEA, KK-IPT evaluated with
  the physical and with the auxiliary hybridization and FEC-AMEA. Both KK-IPT solutions
  invert the magnetization between $U/\Gamma = 4$ and $5$, while AMEA keeps
  $n_{\uparrow} > n_{\downarrow}$ throughout.}
\end{figure}

An exemplary case is discussed below. We observe that for $U > 4\,\Gamma$
plain AMEA and KK-IPT no longer agree on the sign of the magnetization
$m = n_{\uparrow}-n_{\downarrow}$, see Fig.~\ref{fig:breakdown}.

As a matter of fact, while for the fixed hybridization profile, Eq.~\eqref{eq:lorentz},
AMEA keeps $n_{\uparrow} > n_{\downarrow}$ throughout the whole range of
interaction strengths $U$, plain KK-IPT produces a flip in magnetization
occurring at $U > 4\,\Gamma$, regardless of whether the solution
was obtained with the physical or the auxiliary hybridization profile,
see again Fig.~\ref{fig:breakdown}.

However, this does \emph{not} invalidate the correction introduced by the FEC-AMEA
scheme, because only the \emph{ratio} $\Sigma_{\mathrm{IPT}}/\Sigma_{\mathrm{IPT\text{-}aux}}$
enters Eq.~\eqref{eq:mixR}, and since the two KK-IPT solutions exhibit an inverted $m$
at the same $U$, the sign change cancels. This is proof that the fit-error correction strategy proposed
in this work can effectively correct for spurious heating effects even if the individual
impurity solvers do not agree on the overall solution. 

As a matter of fact, FEC-AMEA's magnetization closely tracks the plain AMEA result, deviating by at
most $\sim\!3\%$, with the largest deviation occurring near $U=3\,\Gamma$.

Notably, the difference between the magnetization $m = n_{\uparrow} - n_{\downarrow}$
obtained within KK-IPT for the auxiliary and physical hybridization
is the largest again at $U=3\Gamma$, see again Fig.~\ref{fig:breakdown}.

This difference is {\em not} relevant to the present discussion,
as the extended KK-IPT exhibits the largest difference in $n_{\uparrow} - n_{\downarrow}$
at different values of $U$ when the form of the spin-resolved (physical)
hybridization [Eq.~\eqref{eq:lorentz}] is changed (not shown), which hints
at a non-trivial dependence of the self-consistent determination
of $n_{\sigma}$ and $\mu_{0,\sigma}$ on the external environment
within the KK-IPT scheme. 

\FloatBarrier

\subsection{DMFT for the 2D square lattice}\label{sec:dmft}

Here we discuss the DMFT results for the Hubbard model on a 2D square
lattice with nearest-neighbor hopping (bandwidth $W = 8\,t$) at half filling.

For the characterization of the nonequilibrium properties,
we sandwich the lattice between two metallic contacts
that induce a voltage drop $V = \mu_{\text{L}} - \mu_{\text{R}}$ perpendicular to
the correlated system. Equilibrium conditions are recovered
by setting the leads' chemical potential equal, i.e. $\mu_{\text{L}} = \mu_{\text{R}}$.

For the sake of simplicity, we hereby restrict our analysis to the paramagnetic solution
and postpone the investigation of magnetic systems to future work.
We vary the interaction strength $U$ in order to access both the {\em correlated metal}
and the {\em Mott insulator} regimes, both in equilibrium ($V = 0$) and at finite bias ($V = 1\,t$).

The two metallic leads are characterized by $\mu_{\text{L/R}}=\pm V/2$
and have the same temperature $T = 0.05\,t$, so that the noninteracting distribution
function  $F(\omega)$ at the layer is the average of the two individual
Fermi functions, i.e.\ a two-step function for $V \neq 0$.
The reservoirs are modeled in the  wide-band limit, with a bath hybridisation
$\Gamma_{\mathrm{bath}}$, that sets their retarded component,
i.e. $\gamma^{\text{R}}_{\lambda}(\omega) = - \ii \Gamma_{\mathrm{bath}}$
~\footnote{As usual, the Keldysh components of the metallic contacts are fixed by the fluctuation-dissipation theorem.}.
The hybridisation with the bath is quite small, in order to emphasize the temperature overestimation effects.x
If not stated otherwise, we use $\Gamma_{\mathrm{bath}} = 0.01$ and
$t=1$ as unit of energy.

\begin{figure*}[t]
  \centering
  \includegraphics[width=\textwidth]{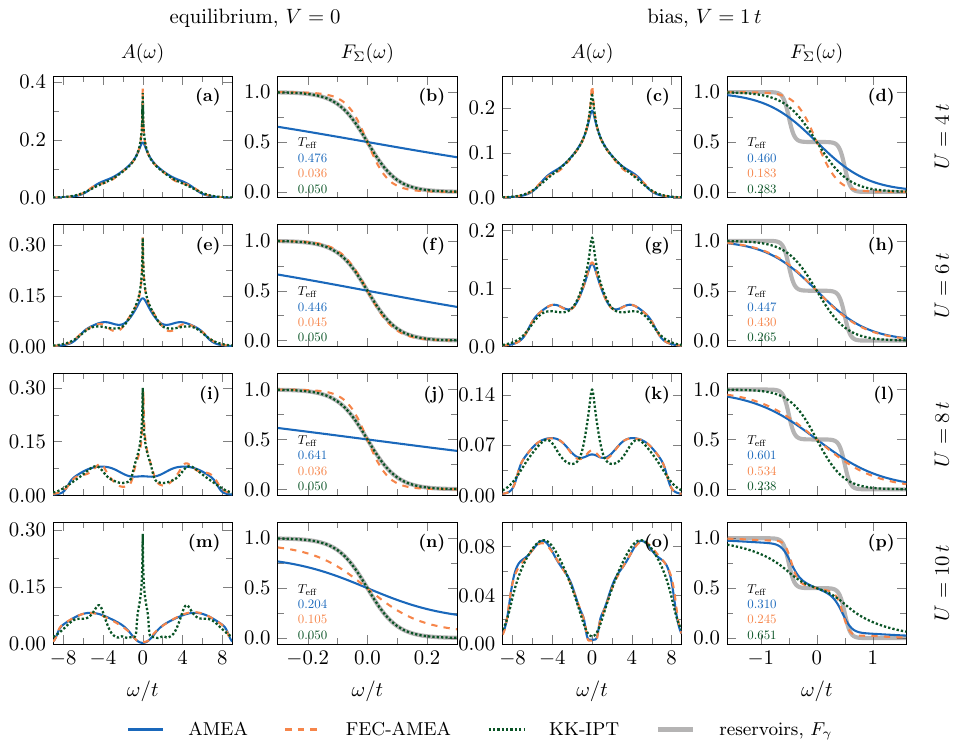}
\caption{\label{fig:dmft_grid} Comparison of the converged DMFT solutions for
  the paramagnetic Hubbard model on the 2D square lattice at half filling,
  coupled to two reservoirs in the wide-band limit with
  $\Gamma_{\mathrm{bath}} = 0.01\,t$ and temperature $T = 0.05\,t$, obtained
  with plain AMEA, KK-IPT, and FEC-AMEA. Rows correspond to different values
  of $U$. The left half shows the equilibrium case, the right half a bias
  voltage $V = t$ between the two reservoirs, $\mu_{\mathrm{L/R}} = \pm V/2$.
  Shown are the local spectral function
  $A(\omega) = -\mathrm{Im}\,G^{R}_{\mathrm{loc}}(\omega)/\pi$,
  Eq.~\eqref{eq:Glocal}, and the distribution function of the SE
  $F_{\Sigma}(\omega)$, Eq.~\eqref{eq:Fdef}, together with the average
  distribution function of the two reservoirs (thick grey line). The values
  of $\Teff$, Eq.~\eqref{eq:Teff}, are given in the panels in units of $t$.
  AMEA uses $N_{B} = 6$ bath sites.}
\end{figure*}

\subsubsection{Equilibrium}\label{sec:dmft:eq}

The left half of Fig.~\ref{fig:dmft_grid} shows the equilibrium DMFT results.

We first discuss the local spectral functions, see Figs.~\ref{fig:dmft_grid}(a), (e), (i) and (m).
At $U = 4\,t$ and $6\,t$ the three schemes yield the same Hubbard bands and
differ only in the QPP at the Fermi level, see Figs.~\ref{fig:dmft_grid}(a)
and (e). In this intermediate coupling regime, where KK-IPT is
reliable, FEC-AMEA and KK-IPT yield practically the same QPP, whereas the one
obtained with plain AMEA is broader and considerably lower.
As will become clear from the distribution functions discussed below, this is
due to the spurious heating effects of plain AMEA rather than to correlations.

At $U = 8\,t$ this difference becomes qualitative, see
Fig.~\ref{fig:dmft_grid}(i). Here, plain AMEA almost completely suppresses the QPP,
with about four times less spectral weight at the Fermi level ($\omega=0$)
than FEC-AMEA, leaving a pseudogap-like dip between the Hubbard bands.
This result alone would suggest that the system is approaching the Mott transition,
whereas FEC-AMEA still describes a correlated metal, as does KK-IPT.

As $U$ is further increased, the situation is reversed, see
Fig.~\ref{fig:dmft_grid}(m) for $U = 10\,t$. In this regime KK-IPT still
retains a sharp QPP and starts to develops spurious satellites next to the
gap, see Fig.~\ref{fig:singleshot}. On the other hand, FEC-AMEA
is in agreement with plain AMEA in that it yields a Mott insulator.
These last two results show
 that FEC-AMEA
follows the solution of plain AMEA, when strong correlation effects become
predominant, while correcting the suppression of the QPP in the correlated
metal due to spurious heating effects present only in plain AMEA.

The advantage of the FEC-AMEA scheme is therefore twofold.
Besides the spectral function, the distribution function is also improved,
this time according to KK-IPT, which inherits the distribution function of its
input hybridization exactly and therefore retains the correct distribution
function in equilibrium.
This is shown in Figs.~\ref{fig:dmft_grid}(b), (f), (j) and (n).
While both KK-IPT and FEC-AMEA reproduce the Fermi function of the reservoirs,
the distribution function obtained with plain AMEA is so smeared out
that within the thermal window ($|\omega| \lesssim T$) it is hardly distinguishable from a straight
line. As a matter of fact, the converged plain AMEA solution corresponds to an
effective temperature about ten times larger than that of the reservoirs,
see the $\Teff$ reported directly on the plots in Figs.~\ref{fig:dmft_grid}(b), (f), (j) and (n).

\begin{figure}[t]
  \centering
  \includegraphics[width=\linewidth]{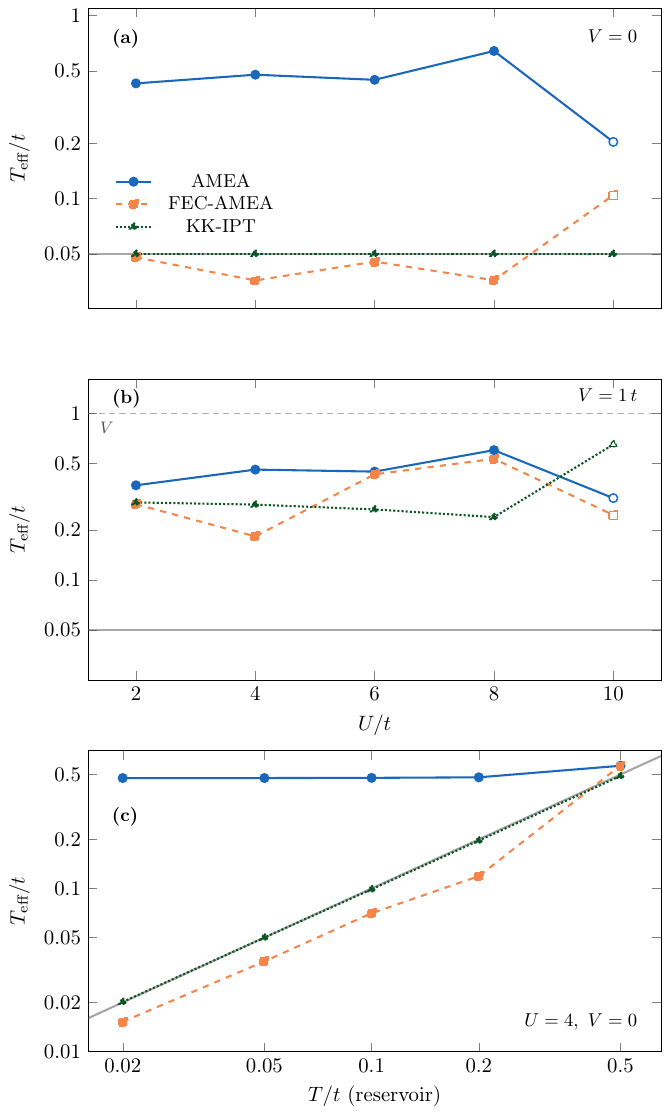}
  \caption{\label{fig:teff_dmft} Effective temperature, Eq.~\eqref{eq:Teff},
  of the converged DMFT solutions of Fig.~\ref{fig:dmft_grid}, obtained with
  plain AMEA, KK-IPT and our FEC-AMEA scheme.
  (a) Equilibrium, as a function of $U$, at the reservoir temperature
  $T = 0.05\,t$ (grey line). (b) Same as (a) at bias $V = 1\,t$ (dashed line).
  Open symbols at $U=10\,t$ denote Mott insulating solutions, for which a temperature
  estimated at the Fermi level ($\omega=0$) is not meaningful.
  (c) $\Teff$ as a function of the reservoir temperature $T$ in equilibrium,
  at fixed $U = 4\,t$. This is the counterpart of Fig.~\ref{fig:teff_ss} for
  the lattice. The grey diagonal denotes $\Teff = T$. Other relevant parameters
  are as in Fig.~\ref{fig:dmft_grid}.}
\end{figure}

Figure~\ref{fig:teff_dmft}(a) shows that this spurious heating happens almost independently
of the value of the interaction strength $U$. Notably, this deviation is much larger than
the one found for the AIM in Sec.~\ref{sec:singleshot}, and it will be further
discussed in Appendix~\ref{sec:dmft:mechanism}.

For KK-IPT, the agreement with the reservoir temperature is expected.
The FEC-AMEA scheme recovers the reservoir temperature as well,
although it slightly underestimates it, by at most $30\,\%$ over the values of $U$
taken into account in this analysis, see again Fig.~\ref{fig:teff_dmft}(a).
This residual error is however about one order of magnitude smaller than the one
introduced by plain AMEA.

Figure~\ref{fig:teff_dmft}(c) shows $\Teff$ as a function of the bath
temperature $T$ at fixed $U = 4\,t$. In particular, the effective temperature
obtained with plain AMEA is essentially independent of $T$, and stays at about
$0.5\,t$ over the whole range of temperatures considered.
The apparent improvement at the highest temperatures is due to the fixed
error becoming small compared to $T$.

On the other hand, KK-IPT follows the diagonal $\Teff = T$ throughout, and
FEC-AMEA runs parallel to it, thus recovering the correct dependence of $\Teff$ on the bath temperature,
although it underestimates $T$ by at most $40\,\%$ over the range of $U$
used in this work.

\subsubsection{Finite bias}\label{sec:dmft:bias}

The nonequilibrium DMFT results for the exemplary case of a
finite bias ($V = 1\,t$) are shown in the right half of Fig.~\ref{fig:dmft_grid},
and in Fig.~\ref{fig:teff_dmft}(b).
In contrast to equilibrium where the self-energy distribution function is expected to equilibrate with the one of the bath, 
there is no exact reference in the nonequilibrium case, so that the error cannot be determined.
The discussions below should therefore be regarded as a first characterization
of the scheme out of equilibrium rather than as a conclusive benchmark.

We start by analysing the local spectral functions shown in
Figs.~\ref{fig:dmft_grid}(c), (g), (k) and (o). At $U = 4\,t$, FEC-AMEA and KK-IPT
yield a similar QPP, which is lower in plain AMEA, see Fig.~\ref{fig:dmft_grid}(c).

At $U = 6$ and $8\,t$, instead, FEC-AMEA follows plain AMEA and yields a
pseudogap-type spectral function, whereas KK-IPT alone retains a pronounced
QPP, see Figs.~\ref{fig:dmft_grid}(g) and (k). At $U = 10\,t$ all three solvers yield
a Mott gap, see Fig.~\ref{fig:dmft_grid}(o).

The distribution functions $F_{\Sigma}$ in Figs.~\ref{fig:dmft_grid}(d), (h), (l) and (p)
retain only a trace of the two-step structure induced by the leads, and are close
to a Fermi-type distribution at a high temperature, which can be attributed
to the redistribution of the injected energy (due to the bias) by electron-electron
scattering.

This observation, accompanied by the absence of a gap, allows us to extract
$\Teff$ also at finite bias, see Fig.~\ref{fig:teff_dmft}(b).
As expected for a bias that is twenty times larger than the bath temperature
and a weak hybridisation ($\Gamma_{\text{bath}} = 0.01\,t$), all three solvers
yield an effective temperature well above that of the leads.

Notably, in the driven case plain AMEA returns an effective temperature
that is close to its overestimated equilibrium value, compare Figs.~\ref{fig:teff_dmft}(a)
and (b). This suggests that the spurious heating is at least comparable to the physical
heating caused by the bias.

At weak coupling, FEC-AMEA reduces the effective temperature considerably: at
$U = 2\,t$ it agrees with KK-IPT, while at $U = 4\,t$ it lies somewhat below it.
At $U = 6$ and $8\,t$, instead, FEC-AMEA yields an effective temperature close
to that of plain AMEA, see again Fig.~\ref{fig:teff_dmft}(b).

We note that the cases in which FEC-AMEA departs from KK-IPT coincide with
those in which plain AMEA and KK-IPT, on their own, converge to qualitatively
different steady states, see Figs.~\ref{fig:dmft_grid}(g) and (k).

A possible interpretation is that the correction introduced by FEC-AMEA,
being of the same magnitude of the error brought about by the fitting procedure,
can mitigate the spurious heating but cannot change the nature of the
self-consistent solution in parameter regimes where the two solvers disagree.
A conclusive answer would require a comparison with a numerically exact
nonequilibrium solver, which is beyond the scope of this work.

The behavior of the distribution functions at finite bias depends on whether
the system is metallic or insulating.
For $U \le 8\,t$ the system is in a driven, correlated metallic state, with a
QPP at $\omega = 0$ resolved in all three schemes, see Figs.~\ref{fig:dmft_grid}(c),
(g) and (k). Hence, there is spectral weight within the bias window
$|\omega| \le V/2$, so that a current can flow from one lead to the other, through the
correlated lattice, and the injected energy is therefore redistributed by electron-electron
scattering.
As a result, $F_{\Sigma}$ does not retain the two-step structure of the leads,
regardless of the solution scheme employed, but rather resembles a Fermi-type
distribution at an increased temperature, see Figs.~\ref{fig:dmft_grid}(d), (h)
and (l).

At $U = 10\,t$, instead, the interaction opens a correlated gap of the order of
the bias, so that the bias window lies almost entirely within the gap and only
barely touches the inner edges of the Hubbard bands, see
Fig.~\ref{fig:dmft_grid}(o). The spectral weight available within the bias
window is more than one order of magnitude smaller than at $U = 8\,t$, and we
therefore expect the current, and with it the energy injected into the system,
to be strongly reduced.
This is reflected in the occupation of the upper Hubbard band, given by
$F_{\Sigma}$ at $\omega \gtrsim 1\,t$, which measures the number of hot
carriers excited by the bias. Compared to $U = 8\,t$, this occupation is
several times smaller for plain AMEA, and about one order of magnitude smaller
for FEC-AMEA, see Figs.~\ref{fig:dmft_grid}(l) and (p), so that the electrons
in the Hubbard bands remain considerably colder than in the metallic case.
KK-IPT is an exception, as it yields a hot distribution also at $U = 10\,t$,
with an effective temperature even higher than in the metallic case, see
Fig.~\ref{fig:teff_dmft}(b).

We stress that, within the gap, the system has essentially no states of its
own, so that $F_{\Sigma}$ closely follows the distribution of the leads,
particularly within FEC-AMEA, see Fig.~\ref{fig:dmft_grid}(p). Hence, the
effective temperature extracted from the slope of $F_{\Sigma}$ at $\omega = 0$,
Eq.~\eqref{eq:Teff}, reflects the leads rather than the electrons in the
Hubbard bands, which is why the Mott insulating solutions are marked by open
symbols in Fig.~\ref{fig:teff_dmft}(b).
A systematic study of this regime, e.g.\ by increasing the bias at
$U = 10\,t$ until the bias window reaches into the Hubbard bands, is left for
future work.

In summary, the FEC-AMEA scheme is most reliable when plain AMEA and KK-IPT
agree on the nature of the solution, which is the case for the metallic
solutions in equilibrium.

Out of equilibrium, we expect it to remain reliable
as long as the bias does not drive the two solvers into qualitatively different phases.

\FloatBarrier

\section{Conclusions}\label{sec:concl}

We have introduced a \fec for AMEA, in which the essentially exact
self-energy of the auxiliary problem is extrapolated to the physical
hybridization function by means of an appropriate ``ratio'' of two self-energies obtained
with an approximate solver, here the nonequilibrium KK-IPT generalized to
spin-polarized solutions.

Our benchmarks lead to three main conclusions. First, at low temperatures the
dominant error of AMEA lies not in the treatment of correlations but in the
distribution function of the bath, and its consequences depend strongly on
the context. For a single impurity, it results in a moderate overestimate of
the temperature, while 
for a Hubbard layer very weakly coupled to thermal reservoirs
 the DMFT self-consistency drives the solution
towards the hottest hybridization function compatible with the lattice, about
an order of magnitude above the reservoir temperature. 
Remarkably, this hot solution has a \emph{smaller}
fit error than the corrected one, so that the quality of the fit alone is not
a reliable indicator of accuracy in a self-consistent calculation.
 Second,
this spurious heating can mimic correlation effects: in the 2D Hubbard model
it suppresses the quasiparticle peak and produces a pseudogap that could be
mistaken for the vicinity of the Mott transition. The \fec removes this
artifact while retaining the Mott insulator at strong coupling, and thus
allows one to separate thermal from correlation effects. Third, the
approximate solver only needs to capture the relative change of the
self-energy between the two hybridization functions: KK-IPT alone misses the
Mott insulator and even inverts the magnetization of the spin-polarized
impurity, yet the ratio entering the correction remains smooth and of order
one.

As a practical consequence, these results suggest that, in finite-bath DMFT calculations
at low temperatures, the effective temperature of the converged solution be
monitored in addition to the fit error. Moreover, the comparison between
plain AMEA and the approximate solver provides a simple consistency check:
the \fec is most reliable when both agree on the nature of the solution, as
for the metallic solutions in equilibrium, whereas a disagreement, as found
at finite bias for intermediate coupling, signals that the corrected result
should be taken with care.

For strong interactions, for which KK-IPT is less reliable, the scheme
would benefit from 
a more accurate approximate strong-coupling solver, for instance the noncrossing
approximation and its extensions~\cite{wi.me.94,ec.we.10,ge.ki.25}.
 Since the construction only requires an
approximate solver that can be applied to arbitrary hybridization functions,
it can be combined with other finite-bath solvers as well, and extended to
multiorbital impurities~\cite{ma.we.25} and to phonons as an additional
dissipation channel~\cite{ma.ga.22,ma.we.23}. Its most immediate application
is to magnetically ordered phases, which survive only well below the ordering
temperature and are therefore particularly sensitive to spurious heating. In
a companion paper we use it to study a correlated altermagnet in equilibrium
and under a perpendicular bias~\cite{fu.ar.26u.sn}, and the same approach
applies to an in-plane electric field~\cite{fu.ar.26u.fd}.

\begin{acknowledgments}
This research was funded by the Austrian Science Fund (FWF) [Grant DOI:10.55776/P33165], and by 
the NaWi Graz. Results have been obtained using the A-Cluster and S-Cluster at TU Graz as well as
the Austrian Scientific Computing (ASC) infrastructure.
\end{acknowledgments}

\section*{Data and Code Availability} 
The data supporting the findings of this study will be openly available in the TU Graz Repository upon publication of this manuscript. 
The code implementing the nonequilibrium iterated perturbation theory impurity solver used in this manuscript can be 
found at \href{https://github.com/mazzocch/neq-kk-ipt-solver}{\texttt{neq-kk-ipt-solver}}. The dataset used to generate the figures in this 
manuscript is available at \href{https://repository.tugraz.at/uploads/pvaa4-8ps02}{\texttt{10.3217/pvaa4-8ps02}}.

\appendix
\section{Discussion about the spurious heating}\label{sec:dmft:mechanism}

The enhanced deviations found within DMFT with respect to the
single-shot solution to the AIM are due to the  feedback
introduced by DMFT itself that amplifies the effects of the fit,
Eq.~\eqref{eq:cost}, within the self-consistent loop.

Since the auxiliary bath cannot represent the sharp Fermi step of
$\Delta^{K}$, the Keldysh component of the SE, $\Sigma^{K}$, obtained with AMEA
is responsible for a slightly higher temperature. Via the lattice Dyson equation,
Eq.~\eqref{eq:Glatt}, the same holds for $G^{K}_{\mathrm{loc}}$, which in turn
determines the new $\Dphys$ via Eq.~\eqref{eq:Delta}, to be fitted in the next
iteration.
As a result, each DMFT iteration increases the temperature of the solution,
until the additional smearing is balanced by the coupling to the reservoirs. On the other hand, the coupling to the reservoir is very weak, so that this balance is shifted to high temperatures.

This is confirmed by the effective temperature of the converged hybridization
function itself, i.e.\ by applying Eq.~\eqref{eq:Teff} to the distribution
function of $\Delta$ instead of $\Sigma$ (not shown).
For plain AMEA, the temperature of the auxiliary hybridization function is already
several times, and up to one order of magnitude, higher than the reservoirs,
whereas for FEC-AMEA it stays close to the bath temperature.
Solving the impurity problem for this hybridization function only adds the
moderate overestimate already found for the AIM in Sec.~\ref{sec:singleshot}.
Hence, the error made by AMEA for a given hybridization function is the same as
for the AIM, but it accumulates over the DMFT iterations.

Further insight into this mechanism is given by the cost function $\chi$,
Eq.~\eqref{eq:cost}. The error $\chi$ [Eq.~\eqref{eq:cost}] is several times
\emph{smaller} for the solution obtained with plain AMEA than that of the
FEC-AMEA scheme. Even though this might sound surprising at first glance,
the effect is explained by the fact that the DMFT loop with plain AMEA
converges to a hybridization function that can be fitted particularly
well with the auxiliary system.

This is due to the finite size of the auxiliary system that effectively smoothens
the physical hybridization, so that DMFT with plain AMEA converges towards
the hottest solution compatible with the lattice.

Within FEC-AMEA this is not the case, and DMFT converges to a more physical, sharper
hybridization function, at the price of a larger fit error, which is however accounted for
by the scheme itself.

\section{Details of the distribution functions for the single impurity case}
\label{sec:distfunc_grid}

\begin{figure*}
  \centering
  \includegraphics[width=\textwidth]{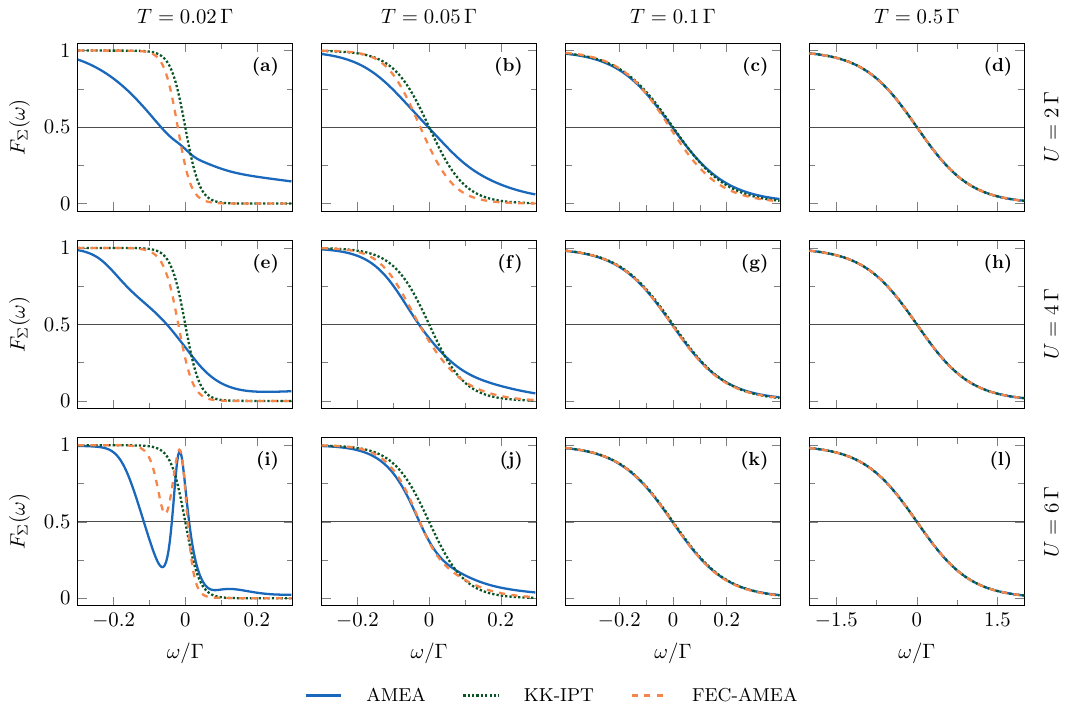}
  \caption{\label{fig:distfunc_grid} Distribution function $F_{\Sigma}(\omega)$,
  Eq.~\eqref{eq:Fdef}, on the same $U = 2, 4, 6\,\Gamma$ and
  $T = 0.02, 0.05, 0.1, 0.5\,\Gamma$ grid as Fig.~\ref{fig:singleshot}, same
  solvers, colors and line styles.  The horizontal line marks
  $F_{\Sigma} = 1/2$. The $\omega$-window is zoomed per column since the
  thermal step narrows with $T$. Other relevant parameters are as in
  Fig.~\ref{fig:singleshot}.}
\end{figure*}

Section~\ref{sec:singleshot} discusses the effective temperature extracted
from $F_{\Sigma}$ (Fig.~\ref{fig:teff_ss}) but never shows $F_{\Sigma}(\omega)$
itself. Figure~\ref{fig:distfunc_grid} does so on the same $(U,T)$-grid as
Fig.~\ref{fig:singleshot}.  For $T \ge 0.1\,\Gamma$ the three solvers coincide
to the eye, consistent with the $\lesssim 4\,\%$ agreement reported there.  At
$T = 0.02$ and $0.05\,\Gamma$ plain AMEA's shallower slope is the smeared,
artificially hot distribution discussed in Sec.~\ref{sec:singleshot}, while
\FEC{} tracks KK-IPT closely. 

Panel (i) in Fig.~\ref{fig:distfunc_grid}, $U = 6\,\Gamma$, $T = 0.02\,\Gamma$,
shows a dip-and-rebound non-monotonicity of the AMEA distribution function,
which the corrected scheme partly inherits. We attribute it to the truncation of
the CI expansion.

In these strongly polarized solutions, the upper Hubbard satellite of the
majority spin carries a spectral weight proportional to the occupation of the
minority species, which can become comparable to the truncation error.
For $N_{\mathrm{exc}} = 3$, this weak satellite is not resolved accurately, and
the spectral function becomes negative in a narrow energy window, i.e.\
$\mathrm{Im}\,G^{R} > 0$ (not shown). At low temperatures, these causality violations are
often accompanied by a distribution function $F_{\Sigma}$ that is not bounded in
the interval $[0,1]$. Raising the number of excitations to
$N_{\mathrm{exc}} = 4$, which is used for all single-impurity results, often removes
these defects in most cases. The non-monotonicity in panel (i) is a residual of
the same truncation error.

We stress that these shortcomings are related to the strongly polarized
impurity considered here. As discussed in Sec.~\ref{sec:singleshot}, the
physical hybridization, Eq.~\eqref{eq:lorentz}, results in a strongly magnetized
impurity. Since the weight of the minority satellite depends on the spin
polarization, we expect the range of interaction strengths $U$ in which these
defects occur to depend on the position and width of the Lorentzian peaks of
the hybridization function. In contrast, the paramagnetic lattice solutions of
Sec.~\ref{sec:dmft} show no causality violation, and we expect magnetic lattice
models with a weaker exchange splitting to be similarly unaffected. In any case,
a causality violation can be detected unambiguously by checking the sign of
$\mathrm{Im}\,G^{R}$ in each run and promptly corrected by increasing $N_{\mathrm{exc}}$.

\section{Spin-dependent nonequilibrium KK-IPT}
\label{sec:spin_kkipt}

The nonequilibrium KK-IPT solver of Ref.~\cite{ma.ar.26} was formulated in the paramagnetic case, $n_{\uparrow} = n_{\downarrow} \equiv n$, for which the onsite energy, hybridization, and occupations do not distinguish the two spin species. Here we show how the same construction generalizes once this restriction is dropped. The generalization follows the equilibrium construction of Ref.~\cite{po.he.98} (built on the atomic-limit argument of Ref.~\cite{ka.ko.96}), carried over to the Keldysh contour along the same lines as Ref.~\cite{ma.ar.26}.

\subsection{Model and Weiss field}

The total Hamiltonian retains the structure $\hat{H} = \hat{H}_{\text{imp}} + \hat{H}_{\text{bath}} + \hat{H}_{\text{hyb}}$ of Ref.~\cite{ma.ar.26}, with each term now spin resolved. The impurity Hamiltonian generalizes to
\begin{equation}\label{eq:spin_h_imp}
\hat{H}_{\text{imp}} = \sum_{\sigma} \varepsilon_{\sigma} \, \hat{f}^{\dagger}_{\sigma} \hat{f}_{\sigma} + U \hat{n}_{\uparrow} \hat{n}_{\downarrow}.
\end{equation}
The bath Hamiltonian reads
\begin{equation}\label{eq:spin_h_bath}
\hat{H}_{\text{bath}} = \sum_{\lambda p \sigma} \left( \varepsilon_{\lambda\sigma} - \mu_{\lambda} \right) \hat{c}^{\dagger}_{\lambda p \sigma} \hat{c}_{\lambda p \sigma},
\end{equation}
where $\lambda = \{\text{l},\text{r}\}$ again labels the left/right lead, $p$ the lead quantum number, and $\varepsilon_{\lambda\sigma}$ allows the lead onsite energy itself to be spin dependent (e.g.\ a Zeeman-split or ferromagnetic reservoir); the lead chemical potential $\mu_{\lambda}$, set by the applied bias as $\mu_{\text{l/r}} = \mu \pm \Phi/2$, is not spin dependent. The impurity-lead coupling generalizes to
\begin{equation}\label{eq:spin_h_hyb}
\hat{H}_{\text{hyb}} = \sum_{\lambda p \sigma} t_{\lambda\sigma} \, \hat{c}^{\dagger}_{\lambda p \sigma} \hat{f}_{\sigma} + \text{H.c.},
\end{equation}
with a possibly spin-dependent hopping amplitude $t_{\lambda\sigma}$ (e.g.\ a spin-polarized contact); no spin-flip tunneling term is included, so the resulting hybridization function stays diagonal in spin.  The latter reads, in direct analogy with Ref.~\cite{ma.ar.26},
\begin{equation}\label{eq:spin_hyb_func}
\underline{\Delta}_{\sigma}(\omega) = \sum_{\lambda} t_{\lambda\sigma}^{2} \, \underline{L}_{\lambda\sigma}(\omega),
\end{equation}
with $\underline{L}_{\lambda\sigma}$ the (possibly spin-dependent) lead GF. The Keldysh structure and the greater/lesser relations of Ref.~\cite{ma.ar.26} carry over unchanged, applied flavor by flavor to every underlined object. Spin dependence of $\underline{\Delta}_{\sigma}$ may thus originate either from $t_{\lambda\sigma}$ or from an intrinsically spin-dependent lead GF $\underline{L}_{\lambda\sigma}$ (e.g.\ a magnetic contact) -- both are supported by the implementation, which accepts either a single (spin-symmetric) shape parameter or a per-flavor dictionary for every hybridization primitive (bandwidth, center, width).

The Weiss field retains the same functional form as the paramagnetic case of Ref.~\cite{ma.ar.26}, applied separately to each flavor with its own auxiliary potential $\mu_{0,\sigma}$,
\begin{equation}\label{eq:spin_weiss_field}
\begin{split}
\mathscr{G}^{\text{R}}_{0,\sigma}(\omega) & = \bigl[ \omega + \mu_{0,\sigma} - \Delta^{\text{R}}_{\sigma}(\omega) \bigr]^{-1}, \\
\mathscr{G}^{\text{K}}_{0,\sigma}(\omega) & = \Delta^{\text{K}}_{\sigma}(\omega) \, \bigl| \mathscr{G}^{\text{R}}_{0,\sigma}(\omega) \bigr|^{2}.
\end{split}
\end{equation}

\subsection{Self-energy ansatz}
\label{sec:spin_se_ansatz}

Because the interaction $U\hat n_{\uparrow}\hat n_{\downarrow}$ only correlates opposite-spin electrons, the Hartree contribution to $\Sigma_{\sigma}$ involves the \emph{opposite}-spin occupation, $Un_{\bar\sigma}$, rather than $Un$ as in the paramagnetic limit (where the two coincide). The same replacement propagates into the second-order (bare-bubble) diagram: in the time domain the diagram entering the KK-IPT ansatz for spin $\sigma$ is
\begin{equation}\label{eq:spin_diagram_time}
\tilde{\Sigma}^{\gtrless}_{\sigma}(\tau) = U^{2} \, \mathscr{G}^{\gtrless}_{0,\sigma}(\tau) \, \mathscr{G}^{\gtrless}_{0,\bar\sigma}(\tau) \, \mathscr{G}^{\lessgtr}_{0,\bar\sigma}(-\tau),
\end{equation}
i.e.\ built from \emph{one} own-spin and \emph{two} opposite-spin Weiss-field propagators rather than three copies of the same propagator. Equation~\eqref{eq:spin_diagram_time} reduces to the paramagnetic diagram of Ref.~\cite{ma.ar.26} once $\sigma = \bar\sigma$, where all three propagators coincide. In the frequency domain, the greater and lesser components read, in direct analogy with the paramagnetic expressions of Ref.~\cite{ma.ar.26},
\begin{equation}\label{eq:spin_lss_gtr_se}
\tilde{\Sigma}_{\sigma}^{\gtrless}(\omega) = U^{2} \int \frac{\dd\omega_{2}}{2\pi} \ I^{\gtrless}_{\sigma}(\omega+\omega_2) \, \mathscr{G}^{\lessgtr}_{0,\bar\sigma}(\omega_{2}),
\end{equation}
\begin{equation}\label{eq:spin_lss_gtr_G0}
I^{\gtrless}_{\sigma}(\omega+\omega_2) \equiv \int \frac{\dd\omega_{1}}{2\pi} \ \mathscr{G}^{\gtrless}_{0,\sigma}(\omega_{1}) \, \mathscr{G}^{\gtrless}_{0,\bar\sigma}(\omega + \omega_{2} - \omega_{1}),
\end{equation}
with $\text{Im}\tilde\Sigma_{\sigma}^{\text{R}}$ and $\tilde\Sigma_{\sigma}^{\text{K}}$ obtained from $\tilde\Sigma_{\sigma}^{\gtrless}$ exactly as in the paramagnetic case of Ref.~\cite{ma.ar.26}, and $\text{Re}\,\tilde\Sigma_{\sigma}^{\text{R}}$ again fixed by the Kramers-Kronig relation.

The spin-dependent KK-IPT self-energy then reads
\begin{equation}\label{eq:spin_se_ansatz}
\begin{split}
\Sigma_{\sigma}^{\text{R}}(\omega) & = U n_{\bar\sigma} + \bigl( 1 - \Sigma^{\text{R}}_{\sigma\beta}(\omega) \bigr)^{-1} \Sigma^{\text{R}}_{\sigma\alpha}(\omega), \\
\Sigma_{\sigma}^{\text{K}}(\omega) & = \bigl( 1 - \Sigma^{\text{R}}_{\sigma\beta}(\omega) \bigr)^{-1} \\
& \times \Bigl[ \Sigma^{\text{K}}_{\sigma\alpha}(\omega) + \Sigma^{\text{K}}_{\sigma\beta}(\omega) \bigl(1 - {\Sigma^{\text{R}}_{\sigma\beta}}^{\ast}(\omega) \bigr)^{-1} {\Sigma^{\text{R}}_{\sigma\alpha}}^{\ast}(\omega) \Bigr],
\end{split}
\end{equation}
with the per-flavor shorthand $\Sigma^{\text{X}}_{\sigma\alpha}(\omega) \equiv \alpha_{\sigma} \tilde\Sigma_{\sigma}^{\text{X}}(\omega)$, $\Sigma^{\text{X}}_{\sigma\beta}(\omega) \equiv \beta_{\sigma} \tilde\Sigma_{\sigma}^{\text{X}}(\omega)$, $\text{X}=\{\text{R},\text{K}\}$ -- here we already anticipate the notation of Sec.~\ref{sec:spin_band_shift} below, see the remark there on why the coefficients are called $\alpha_\sigma,\beta_\sigma$ rather than $A_\sigma,B_\sigma$ as in Ref.~\cite{ma.ar.26}. For fixed $\sigma$, $\Sigma^{\text{X}}_{\sigma\alpha}$ and $\Sigma^{\text{X}}_{\sigma\beta}$ remain proportional to the same $\tilde\Sigma^{\text{X}}_{\sigma}$, so the algebraic argument used in Ref.~\cite{ma.ar.26} to show that the ordering in Eq.~\eqref{eq:spin_se_ansatz} does not matter, and that the resulting Keldysh component is unambiguous, applies exactly flavor by flavor, hence it does not need to be redone here.

\subsection{Interpolation coefficients}
\label{sec:spin_coeffs}

Following the atomic-limit matching of Ref.~\cite{po.he.98} (building on Ref.~\cite{ka.ko.96}), the coefficients generalize to
\begin{equation}\label{eq:spin_ext_ipt_coeffs}
\begin{split}
\alpha_{\sigma} & = \frac{n_{\bar\sigma} \left( 1 - n_{\bar\sigma} \right)}{n_{0,\bar\sigma}\left( 1 - n_{0,\bar\sigma} \right)}, \\
\beta_{\sigma} & = \frac{\left( 1 - n_{\bar\sigma} \right) U + \varepsilon_{\sigma} + \mu_{0,\sigma}}{n_{0,\bar\sigma}\left( 1 - n_{0,\bar\sigma} \right) U^{2}}.
\end{split}
\end{equation}
Both coefficients are fixed by the \emph{opposite}-spin occupations $n_{\bar\sigma}$, $n_{0,\bar\sigma}$: the exact atomic-limit self-energy of species $\sigma$ depends only on $n_{\bar\sigma}$, the only species $\sigma$ interacts with through $U\hat n_{\uparrow}\hat n_{\downarrow}$, while $\beta_{\sigma}$ retains the own-spin onsite energy $\varepsilon_{\sigma}$ and auxiliary potential $\mu_{0,\sigma}$ (through the own-spin Weiss-field pole condition). Setting $\varepsilon_{\uparrow}=\varepsilon_{\downarrow}$ and $\underline\Delta_{\uparrow}=\underline\Delta_{\downarrow}$ makes $\sigma=\bar\sigma$ the unique solution and Eq.~\eqref{eq:spin_ext_ipt_coeffs} reduces to the paramagnetic coefficients of Ref.~\cite{ma.ar.26}, there denoted $A$ and $B$.

We deliberately depart here from the $A,B$ notation of Ref.~\cite{ma.ar.26} and instead write $\alpha_\sigma,\beta_\sigma$, following Ref.~\cite{po.we.97}. The reason is Sec.~\ref{sec:spin_band_shift} below: the optional band-shift refinement introduces two further quantities, conventionally also called $\tilde B$ and $\tilde B_0$, which are effective levels entering the numerator of $\beta_\sigma$ -- and are conceptually unrelated to the interpolation coefficient $\beta_\sigma$ itself. Keeping both under the letter "$B$'' invites exactly the kind of mix-up that letter reuse in physics notation is prone to; using $\alpha_\sigma,\beta_\sigma$ for Eq.~\eqref{eq:spin_ext_ipt_coeffs} throughout keeps the two families of symbols visually and mnemonically distinct.

\subsection{Impurity Green's function and self-consistency}
\label{sec:spin_selfconsistency}

The interacting GF follows the same Dyson form as in Ref.~\cite{ma.ar.26}, applied per flavor,
\begin{equation}\label{eq:spin_imp_gf}
\begin{split}
G_{\sigma}^{\text{R}}(\omega) & = \Bigl[ {\mathscr{G}^{\text{R}}_{0,\sigma}(\omega)}^{-1} - \mu_{0,\sigma} - \varepsilon_{\sigma} - \Sigma_{\sigma}^{\text{R}}(\omega) \Bigr]^{-1}, \\
G_{\sigma}^{\text{K}}(\omega) & = \left( \frac{\mathscr{G}^{\text{K}}_{0,\sigma}(\omega)}{\mid \mathscr{G}^{\text{R}}_{0,\sigma}(\omega) \mid^{2}} + \Sigma_{\sigma}^{\text{K}}(\omega) \right) \mid G_{\sigma}^{\text{R}}(\omega) \mid^{2},
\end{split}
\end{equation}
with occupations $n_{\sigma} = -\ii\int (\dd\omega/2\pi) G_{\sigma}^{<}(\omega)$ and $n_{0,\sigma} = -\ii\int(\dd\omega/2\pi)\mathscr{G}_{0,\sigma}^{<}(\omega)$ defined exactly as in the paramagnetic case.

Unlike the paramagnetic case, the two spin channels do not decouple: Eqs.~\eqref{eq:spin_lss_gtr_se}--\eqref{eq:spin_ext_ipt_coeffs} show that both $\alpha_{\sigma},\beta_{\sigma}$ and the diagram $\tilde\Sigma_{\sigma}$ depend on $\bar\sigma$-flavor quantities. The IPT-$n_{0}$ self-consistency of Ref.~\cite{ma.ar.26} therefore becomes a single, coupled four-dimensional root problem for $\{n_{\sigma},\mu_{0,\sigma}\}_{\sigma=\uparrow,\downarrow}$, generalizing the two paramagnetic conditions $f_1,f_2$ of Ref.~\cite{ma.ar.26} to
\begin{equation}\label{eq:spin_selfconsistency}
n_{\sigma} + \ii \int \frac{\dd\omega}{2\pi} G_{\sigma}^{<}(\omega) = 0, \qquad
n_{0,\sigma} + \ii \int \frac{\dd\omega}{2\pi} \mathscr{G}_{0,\sigma}^{<}(\omega) = 0,
\end{equation}
for $\sigma=\uparrow,\downarrow$, solved simultaneously (four coupled equations in four unknowns) rather than as two independent two-dimensional problems, since a trial value of $n_{\bar\sigma}$ or $n_{0,\bar\sigma}$ feeds directly into the coefficients and diagram of flavor $\sigma$. Setting $\varepsilon_{\uparrow}=\varepsilon_{\downarrow}$ and $\underline\Delta_{\uparrow}=\underline\Delta_{\downarrow}$ reduces Eq.~\eqref{eq:spin_selfconsistency} to the paramagnetic self-consistency of Ref.~\cite{ma.ar.26}, with $n_{\uparrow}=n_{\downarrow}$ the corresponding solution.

\subsection{Spectral moment sum rules}
\label{sec:spin_moments}

The exact first- and second-moment sum rules of Ref.~\cite{ma.ar.26} likewise generalize, spin-resolved, with $n \to n_{\bar\sigma}$ (through the Hartree self-energy) and the hybridization spectral weight taken from the \emph{own}-spin hybridization, $D_{1} \to D_{1,\sigma} \equiv -\pi^{-1}\int \dd\omega\, \text{Im}\Delta_{\sigma}^{\text{R}}(\omega)$:
\begin{equation}\label{eq:spin_first_moment}
M_{1}^{\sigma} = \varepsilon_{\sigma} + U n_{\bar\sigma} = \int \dd\omega \, \omega \, A_{\sigma}(\omega),
\end{equation}
\begin{equation}\label{eq:spin_second_moment}
M_{2}^{\sigma} = \left( M_{1}^{\sigma} \right)^{2} + U^{2} n_{\bar\sigma}(1-n_{\bar\sigma}) + D_{1,\sigma} = \int \dd\omega \, \omega^{2} \, A_{\sigma}(\omega),
\end{equation}
with $A_{\sigma}(\omega) \equiv -\text{Im}G_{\sigma}^{\text{R}}(\omega)/\pi$. As in the paramagnetic case, both moments close on the self-consistent occupations alone (they follow from the exact Hartree limit $\Sigma_{\sigma}^{\text{R}}(\omega\to\infty)\to Un_{\bar\sigma}$ and the atomic-limit second-moment coefficient $U^2 n_{\bar\sigma}(1-n_{\bar\sigma})$, respectively) and thus provide an AMEA-independent internal consistency check of the spin-dependent solution, exactly as they did in the paramagnetic benchmark of Ref.~\cite{ma.ar.26}.

Unlike $M_1^\sigma,M_2^\sigma$, the third moment does not close on the occupations alone, and it is precisely this feature that makes it informative for spin-dependent (magnetic) solutions specifically. Writing $D_{2,\sigma}\equiv-\pi^{-1}\int\dd\omega\,\omega\,\text{Im}\Delta_{\sigma}^{\text{R}}(\omega)$ for the next coefficient of the same large-$\omega$ expansion of $\underline\Delta_\sigma$, the exact third moment reads
\begin{equation}\label{eq:spin_third_moment}
\begin{split}
M_{3}^{\sigma} = {} & \varepsilon_{\sigma}^{3} + U n_{\bar\sigma}\bigl(3\varepsilon_{\sigma}^{2}+3\varepsilon_{\sigma}U+U^{2}\bigr) \\
& + 2 D_{1,\sigma} M_{1}^{\sigma} + D_{2,\sigma} + U^{2} C_{\bar\sigma} = \int \dd\omega \, \omega^{3} \, A_{\sigma}(\omega),
\end{split}
\end{equation}
where the first line is the atomic-limit third moment (the exact result for the isolated, decoupled impurity, recovered when $\underline\Delta_\sigma\to0$) and $2D_{1,\sigma}M_1^\sigma+D_{2,\sigma}$ is the hybridization contribution already fixed by the same $\underline\Delta_\sigma$ entering $M_2^\sigma$. The last term is the two-particle correlator
\begin{equation}\label{eq:spin_third_moment_correlator}
C_{\bar\sigma} \equiv \mathrm{Re}\bigl\langle (2\hat n_\sigma - 1)\hat A^{\dagger}_{\bar\sigma}\bigr\rangle, \qquad \hat A^{\dagger}_{\bar\sigma}\equiv\sum_{\lambda p} t_{\lambda\bar\sigma}\,\hat c^{\dagger}_{\lambda p \bar\sigma}\hat f_{\bar\sigma},
\end{equation}
which is exactly $n_{\bar\sigma}(1-n_{\bar\sigma})\bigl(\tilde B_{\bar\sigma}-\varepsilon_{\bar\sigma}\bigr)$, with $\tilde B_{\bar\sigma}$ the interacting Potthoff-Wegner-Nolting band shift of Eq.~\eqref{eq:spin_band_shift_interacting} below -- i.e.\ $C_{\bar\sigma}$ is precisely the correlator the optional correction of Sec.~\ref{sec:spin_band_shift} is built to reproduce. This is also the origin of the flavor structure worth stressing: the operator $\hat A^\dagger_{\bar\sigma}$ that $C_{\bar\sigma}$ correlates against $(2\hat n_\sigma-1)$ carries only $\bar\sigma$-flavor bath operators, so the $\sigma$-moment genuinely requires a $\bar\sigma$-built quantity -- there is no way to express $M_3^\sigma$ using only same-spin objects, in sharp contrast to $M_1^\sigma,M_2^\sigma$, which (through $n_{\bar\sigma}$ and $D_{1,\sigma}$) still only require one-particle, already-self-consistent data.

Because of this, $M_3^\sigma$ is a genuinely two-particle, cross-flavor test, and correspondingly a much sharper diagnostic than $M_1^\sigma,M_2^\sigma$ for whether a spin-dependent solution captures the correct correlated structure: the plain IPT-$n_0$ scheme of Sec.~\ref{sec:spin_selfconsistency} has no mechanism to enforce it and violates it substantially (by an order-$10\%$ relative deviation already at moderate $U$, growing with coupling, in tests of the reference implementation), whereas the optional band-shift correction of Sec.~\ref{sec:spin_band_shift} satisfies it to the numerical floor by construction, since it is built by self-consistently equating $\tilde B_{\bar\sigma}$ with exactly the functional in Eq.~\eqref{eq:spin_band_shift_interacting}. $M_3^\sigma$ therefore complements $M_1^\sigma,M_2^\sigma$: the first two certify the single-particle sector (occupations, Hartree shift, hybridization weight) regardless of which self-energy scheme is used, while the third specifically probes the spin-resolved two-particle correlations that distinguish the two schemes -- and is consequently the more relevant benchmark for magnetic solutions, where such cross-spin correlations are not incidental but central to the physics.

\subsection{Potthoff-Wegner-Nolting band-shift correction}
\label{sec:spin_band_shift}

Beyond the base ansatz of Secs.~\ref{sec:spin_se_ansatz}--\ref{sec:spin_coeffs}, the solver used in this work can optionally apply the nonequilibrium Keldysh analogue of the Potthoff-Wegner-Nolting~\cite{po.we.97} $m=3$ moment band-shift correction. It enters $\beta_\sigma$ (not $\alpha_\sigma$) as an \emph{additive} correction to the numerator of Eq.~\eqref{eq:spin_ext_ipt_coeffs}, built from two distinct effective levels of the \emph{opposite} spin: a Weiss-field (Hartree-Fock-level) estimate $\tilde B_{0,\bar\sigma}$ and its self-consistent, interacting counterpart $\tilde B_{\bar\sigma}$. Neither should be confused with the coefficient $\beta_\sigma$ itself -- hence the renaming introduced in Sec.~\ref{sec:spin_coeffs}.

The Weiss-field estimate $\tilde B_{0,\bar\sigma}$ is a functional of $\mathscr G_{0,\bar\sigma}$ and $\underline\Delta_{\bar\sigma}$ alone,
\begin{equation}\label{eq:spin_band_shift_weiss}
\tilde B_{0,\bar\sigma} = \varepsilon_{\bar\sigma} + \frac{2n_{0,\sigma}-1}{n_{0,\bar\sigma}(1-n_{0,\bar\sigma})} \left( -\ii \int \frac{\dd\omega}{2\pi} \Bigl[ \mathscr G_{0,\bar\sigma}\Delta_{\bar\sigma} \Bigr]^{<}(\omega) \right),
\end{equation}
where the quantity in square brackets can be expanded by means of the Langreth rules as
\begin{equation}\label{eq:spin_band_shift_langreth_weiss}
 \Bigl[ \mathscr G_{0,\bar\sigma}\Delta_{\bar\sigma} \Bigr]^{<} (\omega) = \mathscr G^{\text{R}}_{0,\bar\sigma}(\omega)\Delta^{<}_{\bar\sigma}(\omega) + \mathscr G^{<}_{0,\bar\sigma}(\omega)\Delta^{\text{A}}_{\bar\sigma}(\omega).
\end{equation}
Its interacting counterpart additionally requires the self-energy and interacting GF of species $\bar\sigma$ at the current trial state,
\begin{equation}\label{eq:spin_band_shift_interacting}
\tilde B_{\bar\sigma} = \varepsilon_{\bar\sigma} + \frac{\displaystyle -\ii \int \frac{\dd\omega}{2\pi} \bigl[ Q_{\bar\sigma} G_{\bar\sigma} \Delta_{\bar\sigma} \bigr]^{<}(\omega)}{n_{\bar\sigma}(1-n_{\bar\sigma})}, \qquad Q_{\bar\sigma} \equiv \frac{2\Sigma_{\bar\sigma}}{U} - 1,
\end{equation}
with the lesser component of the triple product expanded via the Langreth rules,
\begin{equation}\label{eq:spin_band_shift_langreth}
\bigl[Q_{\bar\sigma}G_{\bar\sigma}\Delta_{\bar\sigma}\bigr]^{<} = Q^{\text{R}}_{\bar\sigma}G^{\text{R}}_{\bar\sigma}\Delta^{<}_{\bar\sigma} + Q^{\text{R}}_{\bar\sigma}G^{<}_{\bar\sigma}\Delta^{\text{A}}_{\bar\sigma} + Q^{<}_{\bar\sigma}G^{\text{A}}_{\bar\sigma}\Delta^{\text{A}}_{\bar\sigma},
\end{equation}
where $Q^{\text{R}}_{\bar\sigma}=2\Sigma^{\text{R}}_{\bar\sigma}/U-1$ and $Q^{<}_{\bar\sigma}=2\Sigma^{<}_{\bar\sigma}/U$. The correction then enters $\beta_\sigma$ additively,
\begin{equation}\label{eq:spin_beta_corrected}
\beta_{\sigma} = \frac{\left( 1 - n_{\bar\sigma} \right) U + \varepsilon_{\sigma} + \mu_{0,\sigma} + \bigl( \tilde B_{\bar\sigma} - \tilde B_{0,\bar\sigma} \bigr)}{n_{0,\bar\sigma}\left( 1 - n_{0,\bar\sigma} \right) U^{2}},
\end{equation}
replacing Eq.~\eqref{eq:spin_ext_ipt_coeffs}. Note that $\varepsilon_{\bar\sigma}$ never itself appears in the base $\beta_\sigma$ of Eq.~\eqref{eq:spin_ext_ipt_coeffs} (only the own-spin level $\varepsilon_{\sigma}$ does): the correction is a genuinely new additive term, not a substitution of one already present. It vanishes, and Eq.~\eqref{eq:spin_ext_ipt_coeffs} is recovered exactly, whenever $\tilde B_{\bar\sigma}=\tilde B_{0,\bar\sigma}$ -- in particular at the start of the inner iteration described next, where the interacting estimate is seeded from the Weiss-field one.

Because Eq.~\eqref{eq:spin_band_shift_interacting} requires $\Sigma_{\bar\sigma}$ and $G_{\bar\sigma}$, which themselves depend on $\beta_{\bar\sigma}$ through Eq.~\eqref{eq:spin_beta_corrected}, the correction must be found by fixed-point iteration \emph{inside} every trial state visited by the outer root solve of Sec.~\ref{sec:spin_selfconsistency}, not once per converged solution. The pseudocode below makes explicit where each quantity is computed, matching the nested-loop structure of the reference implementation.

\begin{widetext}
\noindent\textbf{Algorithm~1.} One evaluation of the spin-dependent nonequilibrium KK-IPT residual, at trial point $(\mu_{0,\sigma},n_{\sigma})_{\sigma=\uparrow,\downarrow}$; called repeatedly by the outer four-dimensional root solve for Eq.~\eqref{eq:spin_selfconsistency}.
\begin{algorithmic}[1]
\For{$\sigma \in \{\uparrow,\downarrow\}$}
  \State $\mathscr G_{0,\sigma} \gets$ Eq.~\eqref{eq:spin_weiss_field} from $\mu_{0,\sigma}$, $\underline\Delta_\sigma$; \quad $n_{0,\sigma} \gets -\ii \int \frac{\dd\omega}{2\pi}\,\mathscr G_{0,\sigma}^{<}(\omega)$
\EndFor
\For{$\sigma \in \{\uparrow,\downarrow\}$}
  \State $\tilde\Sigma_{\sigma}^{\gtrless} \gets$ Eqs.~\eqref{eq:spin_lss_gtr_se}--\eqref{eq:spin_lss_gtr_G0} from $\mathscr G_{0,\sigma},\mathscr G_{0,\bar\sigma}$ \Comment{depends only on $\mathscr G_0$: computed once, reused below}
\EndFor
\If{Potthoff correction \textbf{enabled}}
  \For{$\sigma \in \{\uparrow,\downarrow\}$}
    \State $\tilde B_{0,\sigma} \gets$ Eq.~\eqref{eq:spin_band_shift_weiss}
  \EndFor
  \State $\tilde B_\sigma \gets \tilde B_{0,\sigma}$ for $\sigma \in \{\uparrow,\downarrow\}$ \Comment{seed: $\beta_\sigma$ starts out equal to Eq.~\eqref{eq:spin_ext_ipt_coeffs}}
  \Repeat
    \For{$\sigma \in \{\uparrow,\downarrow\}$}
      \State $\beta_\sigma \gets$ Eq.~\eqref{eq:spin_beta_corrected} using the \emph{current} $\tilde B_{\bar\sigma},\tilde B_{0,\bar\sigma}$; \quad $\Sigma_\sigma \gets$ Eq.~\eqref{eq:spin_se_ansatz}
    \EndFor
    \For{$\sigma \in \{\uparrow,\downarrow\}$}
      \State $G_\sigma \gets$ Eq.~\eqref{eq:spin_imp_gf}
    \EndFor
    \For{$\sigma \in \{\uparrow,\downarrow\}$}
      \State $\tilde B_\sigma^{\text{new}} \gets$ Eq.~\eqref{eq:spin_band_shift_interacting} from $\Sigma_\sigma,G_\sigma,\underline\Delta_\sigma,n_\sigma$
    \EndFor
    \State $\tilde B_\sigma \gets (1-\eta)\,\tilde B_\sigma + \eta\,\tilde B_\sigma^{\text{new}}$ for $\sigma \in \{\uparrow,\downarrow\}$ \Comment{linear mixing, $\eta\in(0,1]$}
  \Until{$\max_\sigma |\tilde B_\sigma^{\text{new}} - \tilde B_\sigma|$ below tolerance, or iteration budget exhausted}
  \State \emph{rebuild} $\Sigma_\sigma,G_\sigma$ once more via Eqs.~\eqref{eq:spin_se_ansatz},\eqref{eq:spin_imp_gf} using the final $\tilde B_\sigma$ \Comment{the last pass above used the pre-mixing $\tilde B_\sigma$}
\Else
  \For{$\sigma \in \{\uparrow,\downarrow\}$}
    \State $\beta_\sigma \gets$ Eq.~\eqref{eq:spin_ext_ipt_coeffs}; \quad $\Sigma_\sigma \gets$ Eq.~\eqref{eq:spin_se_ansatz}; \quad $G_\sigma \gets$ Eq.~\eqref{eq:spin_imp_gf}
  \EndFor
\EndIf
\State \Return $\Bigl( n_\sigma + \ii\!\int\!\frac{\dd\omega}{2\pi}G_\sigma^{<}(\omega),\ \ n_\sigma + \ii\!\int\!\frac{\dd\omega}{2\pi}\mathscr G_{0,\sigma}^{<}(\omega) \Bigr)_{\sigma=\uparrow,\downarrow}$ \Comment{Eq.~\eqref{eq:spin_selfconsistency}}
\end{algorithmic}
\end{widetext}

The correction restores the exact third spectral moment $M_3^\sigma$, whose closed form involves the two-particle correlator $C_{\bar\sigma}\equiv n_{\bar\sigma}(1-n_{\bar\sigma})\bigl(\tilde B_{\bar\sigma}-\varepsilon_{\bar\sigma}\bigr)$ [cf. Eq.~\eqref{eq:spin_band_shift_interacting}], beyond the accuracy of the plain IPT-$n_0$ scheme of Sec.~\ref{sec:spin_selfconsistency}. It is included here as an available refinement of the spin-dependent solver, not as part of the base ansatz benchmarked in the main text.

\section{Sum rules}\label{app:sumrules}

Reference~\cite{ma.ar.26} derives the exact first- and second-moment spectral
sum rules of the single-band Anderson impurity model,
\begin{equation}
  M_{1,\sigma} = \varepsilon_{\sigma} + U n_{\bar\sigma}
   = \int \dd\omega\,\omega\,A_{\sigma}(\omega) ,
  \label{eq:M1}
\end{equation}
\begin{equation}
  M_{2,\sigma} = M_{1,\sigma}^{2} + U^{2} n_{\bar\sigma}(1-n_{\bar\sigma}) + D_{1,\sigma}
   = \int \dd\omega\,\omega^{2}\,A_{\sigma}(\omega) ,
  \label{eq:M2}
\end{equation}
with $D_{1,\sigma} \equiv -\pi^{-1}\int\dd\omega\,\mathrm{Im}\,\Delta^{\mathrm
R}_{\sigma}(\omega)$ the zeroth moment of the hybridization, and shows that
KK-IPT satisfies both to better than $0.05\,\%$ by construction of the
ansatz.  Since Eq.~\eqref{eq:mixR} rescales
$\mathrm{Im}\,\Sigma^{\mathrm{R}}_{\mathrm{AMEA}}$ by a ratio of two such
Ans\"atze rather than being itself fit to the moments, the sum rules are not
manifestly preserved by the mixing.

The first moment is nevertheless exact for $\Sigma_{\mathrm{FEC}}$ by
construction: Eq.~\eqref{eq:mixKK} adds the exact Hartree term
$Un_{\bar\sigma}$ directly, so $\mathrm{Re}\,\Sigma^{\mathrm
R}_{\mathrm{FEC},\sigma}(\omega\to\infty)\to Un_{\bar\sigma}$ regardless of
the mixing ratio.  We checked both sum rules explicitly against the
single-shot benchmark of Sec.~\ref{sec:singleshot}, where $\Delta^{\mathrm
R}_{\mathrm{ph}}$ is known analytically [Eq.~\eqref{eq:lorentz}, giving
$D_{1,\sigma}=\Gamma$ exactly] and $A_{\sigma}(\omega)$ can be rebuilt
directly from the stored $\Sigma_{\mathrm{FEC}}$ via the Dyson equation.
Across the full $U\in\{2,4,6,8\}\,\Gamma$, $T\in\{0.02,0.05,0.1,0.5\}\,\Gamma$
grid and both spins, $M_{1}$ is reproduced to better than $0.03\,\%$ --
consistent with being exact up to the finite frequency grid -- and $M_{2}$ to
better than $2.8\,\%$, improving with $U$ (below $0.6\,\%$ for $U\ge4\,\Gamma$,
the regime used in the DMFT benchmark of Sec.~\ref{sec:dmft}).  The reason is
that the corrected self-energy's own large-$\omega$ expansion carries the
exact atomic-limit coefficient $U^{2}n_{\bar\sigma}(1-n_{\bar\sigma})$,
evaluated at the fit-error corrected solution's own occupation, as its subleading term, to
essentially the same accuracy as the Hartree term -- even though neither
Eq.~\eqref{eq:mixR} nor Eq.~\eqref{eq:mixKK} enforces this explicitly.

\nocite{ra.sm.86}
\bibliography{references_database,my_refs}

\end{document}